%% file: main.tex
\documentclass[sigconf, screen]{acmart}
\usepackage{ifthen}

\usepackage{balance}
\usepackage{multirow}
\usepackage{pifont}
\usepackage{semantic}
\usepackage{framed}
\usepackage[ruled,linesnumbered,vlined]{algorithm2e}
\usepackage{ulem}

\renewcommand{\em}{\it}
\renewcommand{\emph}[1]{\textit{#1}}

{\endMakeFramed}

\newif\ifdiff
\difftrue
\difffalse
\ifdiff
\newcommand{\revise}[1]{\textcolor{blue}{#1}}
\newcommand{\delete}[1]{\textcolor{red}{\sout{#1}}}
\newcommand{\replace}[2]{\textcolor{red}{\sout{#1}}\textcolor{blue}{#2}}
\newcommand{\startreviseblock}{\color{blue}}
\newcommand{\dismissreviseblock}{\color{black}}
\newcommand{\remark}[1]{\textcolor{orange}{[remark: #1]}}
\else
\newcommand{\revise}[1]{{{#1}}}
\newcommand{\delete}[1]{}
\newcommand{\replace}[2]{#2}
\newcommand{\startreviseblock}{}
\newcommand{\dismissreviseblock}{}
\newcommand{\remark}[1]{}
\fi

\newif\ifshowminordiff
\showminordifftrue
\showminordifffalse
\ifshowminordiff
\newcommand{\mrevise}[1]{\revise{#1}}
\newcommand{\mdelete}[1]{\delete{#1}}
\newcommand{\mreplace}[2]{\textcolor{red}{\sout{#1}}\textcolor{blue}{#2}}
\newcommand{\mremark}[1]{\remark{#1}}

\else
\newcommand{\mrevise}[1]{#1}
\newcommand{\mdelete}[1]{}
\newcommand{\mreplace}[2]{#2}
\newcommand{\mremark}[1]{}

\fi

\setcopyright{none} 
\acmConference[Anonymous Submission to ACM CCS 2022]{ACM Conference on Computer and Communications Security}{Due 2 May 2022}{Los Angeles, USA}

\newcommand{\tool}{{Netlifter}}
\newcommand{\defpar}[1]{\smallskip\noindent\textbf{#1}}
\newcommand{\ffbox}[1]{\smallskip\noindent\fbox{#1}\smallskip}
\newcommand{\fieldnamefun}[1]{\textsf{name}(#1)}
\newcommand{\fieldname}[1]{\textsf{`#1'}}
\newcommand{\varname}[1]{\textsf{#1}}
\newcommand{\smashit}[1]{\textit{\smash{#1}}}
\newcommand{\usmashit}[1]{\underline{\textit{\smash{#1}}}}

\begin{document}

\title[Lifting Network Protocol Implementation to Precise Format Specification]{Lifting Network Protocol Implementation to Precise Format Specification with Security Applications}

\author{Qingkai Shi}
\affiliation{%
    \institution{Purdue University}
    \country{USA}
}
\email{shi553@purdue.edu}

\author{Junyang Shao}
\affiliation{%
    \institution{Purdue University}
\country{USA}
}
\email{shao156@purdue.edu}

\author{Yapeng Ye}
\affiliation{%
    \institution{Purdue University}
\country{USA}
}
\email{ye203@purdue.edu}

\author{Mingwei Zheng}
\affiliation{%
    \institution{Purdue University}
\country{USA}
}
\email{zheng618@purdue.edu}

\author{Xiangyu Zhang}
\affiliation{%
    \institution{Purdue University}
    \country{USA}
}
\email{xyzhang@cs.purdue.edu}

\input{abstract}



%\keywords{Reverse engineering; abstract interpretation; graph rewriting.}

\maketitle

\input{introduction}

\input{motivation}

\input{overview}

\input{approach}

\input{evaluation}

\input{relwork}

\input{conclusion}

\balance
\bibliographystyle{ACM-Reference-Format}
\bibliography{sigproc}

\input{appendix}

\end{document}

%% file: abstract.tex
\begin{abstract}
    Inferring protocol formats is critical for many security applications.
    However,
    existing format-inference techniques often miss many formats,
    because almost all of them are in a fashion of dynamic analysis
    and rely on a limited number of network packets to drive their analysis.
    If a feature is not present in the input packets, the feature will be missed in the resulting formats.
    We develop a novel static program analysis for format inference.
    It is well-known that static analysis does not rely on any input packets
    and can achieve high coverage by scanning every piece of code.
    However,
    for efficiency and precision,
    we have to address two challenges, namely path explosion and disordered path constraints.
    To this end,
    our approach uses abstract interpretation to produce a novel data structure called the abstract format graph.
    It delimits precise but costly operations to only small regions,
    thus ensuring precision and efficiency at the same time.
    Our inferred formats are of high coverage and precisely specify both field boundaries
    and semantic constraints among packet fields.
    Our evaluation shows that we can infer formats for a protocol in one minute with >95\% precision and recall, much better than four baseline techniques. 
    Our inferred formats can substantially enhance existing protocol fuzzers,
    improving the coverage by 20\% to 260\% and discovering 53 zero-days with 47 assigned CVEs.
    We also provide case studies of adopting our inferred formats in other security applications including traffic auditing and intrusion detection.
\end{abstract}

%% file: introduction.tex
\section{Introduction}
\label{sec:intro}

%Computing systems are becoming increasingly interconnected with emerging paradigms such as the internet of things. 
Network protocols define how computing systems are connected.
Thus, security vulnerabilities in network protocols may have devastating consequences.
For example, the WannaCry attack, which was caused by a protocol vulnerability, led to over \$8 billion loss across 150 countries~\cite{ibmxforce2018report}.
To aid automated security analysis for network protocols,
a formal specification of packet formats is often mandatory
--- it enables security testing by facilitating the generation of legitimate network packets for security testing~\cite{gascon2015pulsar,oehlert2005violating}; they are the foundations for protocol model checking~\cite{musuvathi2004model,bishop2005rigorous} and formal verification~\cite{bishop2006engineering}; and they can guide automated code generation with strong guarantees~\cite{wang2022profactory}.

However,
while protocols may have their specification documents in natural languages,
formal, or machine-readable, packet formats are often not available, and even when they are,
they may be incomplete or inaccurate~\cite{bastani2017synthesizing}.
Therefore, automatically inferring formal protocol formats is of importance. There are three typical scenarios. {First}, the protocol implementation is not accessible but network packets are available. In this case, network trace analysis~\cite{cui2007discoverer,kleber2018nemesys,kleber2020message,ye2021netplier,wang2011biprominer,wang2012semantics,luo2013position} are proposed. They leverage statistical analysis and machine learning to infer how a packet can be divided into fields.
Since the underlying techniques have inherent uncertainty,
the quality of inferred formats tends to be insufficient to drive many security applications such as protocol fuzzing.
We call them the \underline{\smashit{category-one}} techniques.

In the {second} scenario, the executable code of a protocol and a set of valid packets are available. Dynamic program analysis~\cite{gopinath2020mining,hoschele2016mining,caballero2009dispatcher,lin2008automatic,cui2008tupni,caballero2007polyglot,wang2009reformat,lin2010reverse,liu2013extracting} then trace how individual packet bytes are propagated when running the code on the provided packets. The protocol formats then can be inferred from the data/control flow relations collected at runtime. For example, a typical rule to infer a raw data field is that consecutive bytes in the field are accessed by the same instruction~\cite{
caballero2009dispatcher,lin2008automatic,cui2008tupni}. These techniques can precisely infer the syntax of provided packets and denote the state-of-the-art.
In some cases (e.g.,~\cite{cui2008tupni}), semantics constraints, for example, those describing correlations across packet fields like the sum of fields $A$ and $B$ cannot exceed a certain threshold, can be inferred as well. 
However, the inferred formats are often incomplete when the provided packets do not have good coverage of all possible formats. 
We call them the \underline{\smashit{category-two}} techniques.

We focus on the {third} scenario, in which the source code of a network protocol is available. 
%This is a very common scenario, in which most open-source protocols fall. 
%Note that it is wrong to assume that source code alone is sufficient for most security applications. 
As we will show in Section~\ref{sec:eval}, open-source protocol implementations have many zero-day vulnerabilities. 
Without precise formal formats, existing protocol fuzzers such as BooFuzz~\cite{boofuzz} can hardly find them.
In a recent work FRAME-SHIFTER~\cite{jabiyev2022frameshifter}, critical bugs were found by fuzzing HTTP/1 and HTTP/2, whose implementations are publicly available. While the authors manually crafted the protocol formats, automatic format inference can generalize their method to other protocols.
In network traffic auditing and attack detection, e.g., using Wireshark~\cite{wireshark} and Snort~\cite{snort}, substantial manual efforts are still needed to write dedicated protocol parsers for Wireshark and Snort
even when a protocol is open-sourced. In contrast, with our inferred formal formats,
Wireshark and Snort can be easily extended.

In this paper, we focus on the third scenario and develop a static program analysis to produce formal protocol formats, including both syntax and semantics, from the source code. We call it a {protocol lifting} technique, belonging to \underline{\smashit{category-three}}.
We resort to static analysis in order to address the coverage problem in dynamic analysis.
Meanwhile, high accuracy can be achieved as it adopts a path-sensitive analysis. 
We produce BNF-like protocol formats.
While BNF~\cite{crocker1997augmented} is a common language to describe syntax, we enhance it to include first-order-logic semantic constraints across protocol fields. 
%We choose to use BNF-like formats as they align well with how humans read/describe protocols in natural languages and, meanwhile, allow us to further generate protocol formats/parsers in other languages such as C/C++ and P~\cite{desai2013p}.
As we will show in \S\ref{sec:nutshell}, lifting source code to protocol formats is highly challenging. 
First, the traditional data-flow analysis that 
aggregates analysis results of multiple program paths at their joint point yields very poor results, whereas path-sensitive analysis that considers individual paths separately is prohibitively expensive due to path explosion. Second, 
the inferred formats are mostly out of order for human interpretation, which is highly undesirable as humans are important consumers of the formats in security applications.

To address the challenges, we develop a novel static analysis. In particular, we develop abstract interpretation rules that can derive an {abstract format graph} (AFG) from the source code.
AFG can be considered as a transformed control flow graph.
It precludes statements that are irrelevant to packet formats.
It further merges program subpaths that are irrelevant to formats so that path-sensitive analysis is not performed on the merged places.
Meanwhile, it retains sufficient information such that a localized but precise path-sensitive analysis can be performed on the unmerged parts of the graph.
Therefore, it mitigates the path-explosion problem without losing analysis accuracy.
The AFG is further unfolded and reordered to generate BNF-style production rules and first-order-logic formulas that describe semantic constraints across protocol fields.
In summary, we make the following four contributions:
\begin{itemize} %[\topsep=1.5pt]
    \item We develop an abstract interpretation method that produces a novel representation, namely the abstract format graph,
    to facilitate format inference.\smallskip

    \item We propose a localized graph unfolding algorithm that can perform precise path-sensitive analysis in small AFG regions to significantly mitigate path explosion.\smallskip
    
    \item We devise a graph reordering algorithm that translates an unfolded AFG to the commonly-used BNF so that our inferred formats can be widely applied in practice.\smallskip

    \item 
    We implement our approach as a tool, namely \tool, \revise{to infer packet formats from protocol parsers written in C. We}
    \delete{and }evaluate it on a number of protocols from different domains.
    \tool\ is highly efficient as it can infer formats in one minute.
    \tool\ is highly precise with a high recall
    as its inferred formats uncover $\ge 95\%$ formats with $\le 5\%$
    false ones. 
    In contrast, the baselines, 
    often miss >$50\%$ of formats and, sometimes, produce >$50\%$ false ones.
    We use the inferred formats to enhance grammar-based protocol fuzzers,
    which are improved by 20\%-260\% in terms of coverage
    and detect 53 zero-day vulnerabilities with 47 assigned CVEs.
    Without our formats, only 12 can be found.
    We also provide case studies of adopting our approach in traffic analysis and intrusion detection.
    \tool\ is publicly available~\cite{netlifter}.
    %We will make \tool\ available upon publication.
\end{itemize}

\newpage

%% file: motivation.tex
\section{Motivation}
\label{sec:motivation}

\begin{figure}
    \includegraphics[width=\columnwidth]{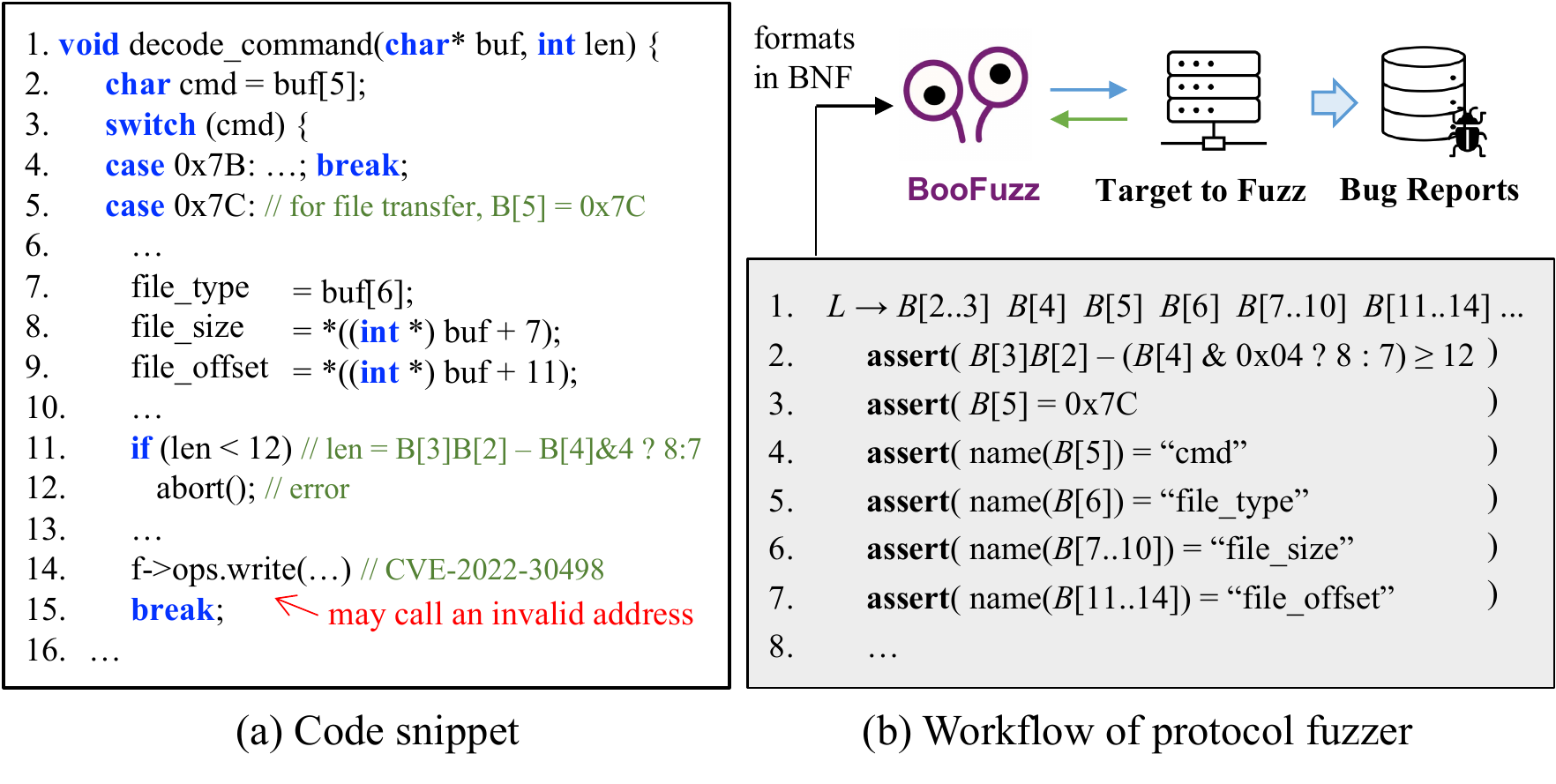}
    \caption{(a) Simplified code that parses the file-transfer command. (b) The typical workflow of protocol fuzzers with a snippet of the format inferred by \tool, in which the first row is a BNF production rule denoting syntax (e.g., field partitioning) and the remaining denote semantic constraints.}
    \label{fig:motivation-1}
\end{figure}

We use an open-source protocol, namely {Open Supervised Device Protocol} (OSDP), to illustrate the limitations of existing methods and how our technique can facilitate various security applications. OSDP is an access control communications standard developed by the {Security Industry Association} to improve interoperability among access control and security products. 
%It was released in December 2020 and has become increasingly popular in government and other higher-security settings. 
Although it is an open-source protocol, its full specification is not publicly available.
The only available document~\cite{libosdp2022doc} lacks many details.
For instance, it includes the formats for only 7 out of the 27 supported~commands.

%\smallskip
%\noindent
%{\bf Exploiting OSDP and Defense Insufficiency.} 
%Although OSDP is designed for secure access control, 
The implementation of OSDP is vulnerable. 
Figure~\ref{fig:motivation-1}(a) shows a code snippet related to a zero-day bug found by 
a protocol fuzzer enhanced by our approach. 
%We reported the bug to the developers and they fixed it immediately.
%Briefly,
The code shows part of the packet~parsing function.
The variable \textsf{buf} is a byte array representing the OSDP packet,
and we use $B[i]$ to represent the ($i+1$)th byte in the packet.
The bug is at Line 14. It may invoke an invalid function pointer \textit{f->ops.write}, which could lead to a crash or be exploited for DoS or ROP attacks.
%Assume an attacker has found the bug and crafts an input to hijack an IoT device. He leverages the device to launch an attack on a government server.
There are multiple ways to 
%improve the security of OSDP and 
avoid such attacks. The first one is to use fuzzing techniques to find bugs in its implementation and have them fixed before exploitation. The second is to provide OSDP support in network traffic analysis and attack detection tools such that the attack can be analyzed and further prevented. However, existing methods fall short as discussed below.

\defpar{Standard Network Fuzzing Can Hardly Find the Zero-day.}
Different from stand-alone application fuzzers such as AFL~\cite{afl}, network fuzzers, such as BooFuzz~\cite{boofuzz}, often operate in a client-server architecture.
The server runs the target protocol implementation. The client leverages grammar-based fuzzing to generate packets as per the formats, send the packets to the server, receive responses, and generate new packets to fuzz the target.
However, the effectiveness of these fuzzers hinges on the protocol formats. When the formats are not available like in our OSDP case, 
they quickly degenerate into traditional greybox fuzzers that arbitrarily mutate bits or bytes. Such mutated packets can hardly pass many input validity checks in the code. 
For example, in Figure~\ref{fig:motivation-1}(a), to expose the bug, a fuzzer has to get through the check at Line 11, which is a complex relation across multiple fields as shown in the comment.
%There are many other checks similar to Line 11.
As a result, standard network fuzzers fail to find the bug when an imprecise or incomplete format of OSDP is provided.

\newpage

\begin{figure}[t]
    \includegraphics[width=\columnwidth]{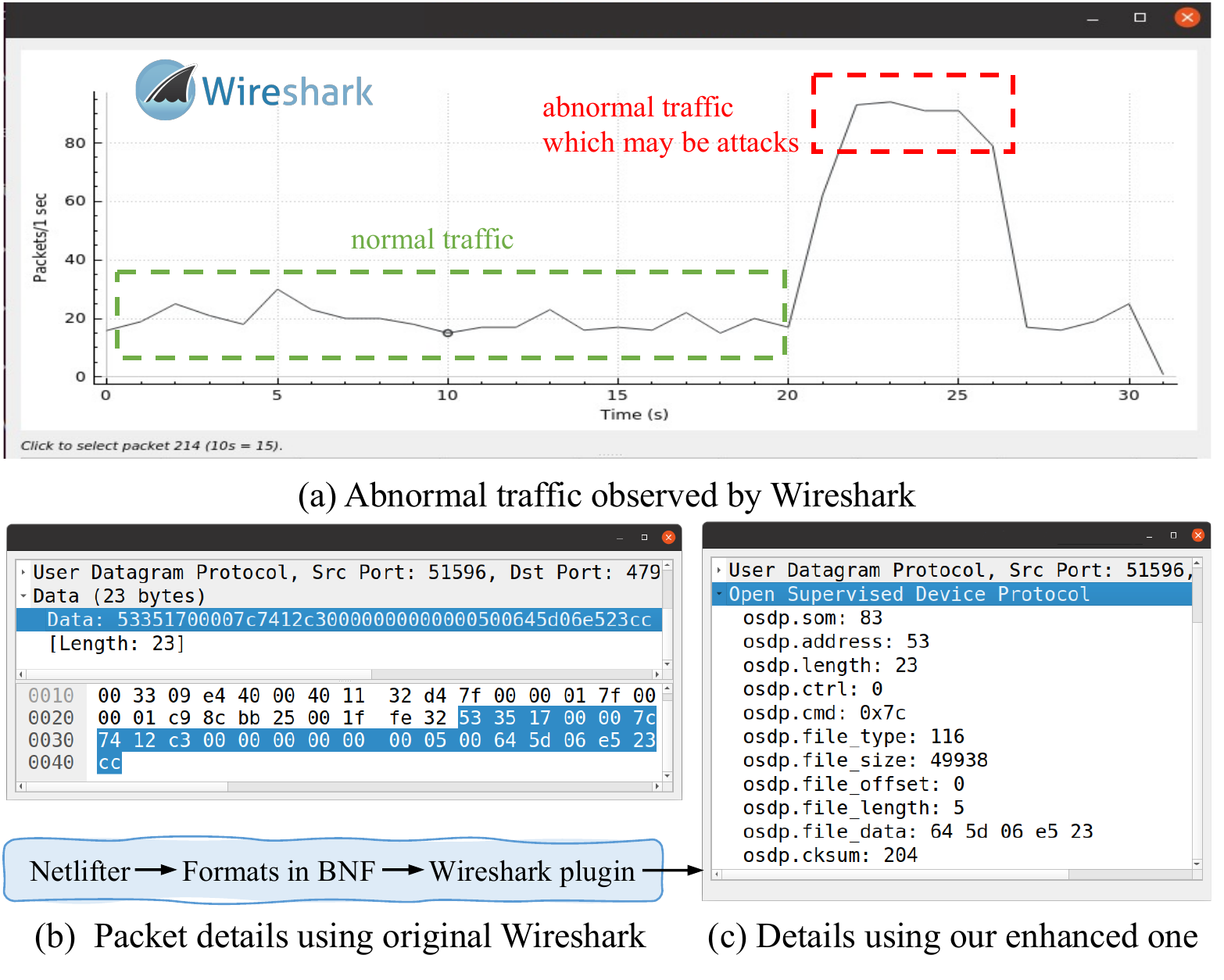}
    \caption{Network traffic auditing via Wireshark.}
    \label{fig:motivation-2}
\end{figure}

\defpar{Lack of Support for OSDP in Wireshark and Snort.}
We can also rely on network traffic analyzers, e.g., Wireshark~\cite{wireshark}, 
and attack detection tools, e.g., Snort~\cite{snort}, to ensure security.
However,
both Wireshark and Snort do not support OSDP. 
Assume
Wireshark is deployed at the gateway.
It detects abnormal traffic as highlighted in the red box in Figure~\ref{fig:motivation-2}(a). Note that in the diagram the x-axis is time and the y-axis denotes the amount of traffic per second.
However, the traffic is not interpretable for Wireshark as OSDP is not supported. Instead, the OSDP packets are treated as raw data bytes as shown in Figure~\ref{fig:motivation-2}(b). Thus, it is hard to analyze the packet details and determine which device launches the attack.

\subsection{Limitations of Existing Techniques}

A way to address the aforestated defense insufficiency is to infer the protocol formats.
As discussed in \S\ref{sec:intro}, existing techniques fall into categories one and two. Category-one infers formats from a set of network packets.
For example, a recent method NetPlier~\cite{ye2021netplier} leverages a probabilistic analysis on network packets to determine
a keyword field, i.e., the field identifying the packet type, by computing the probabilities of each byte offset. Once the keyword field is determined, it clusters
packets according to the value of the field and
applies multi-sequence alignment to derive message format. However, real-world network packets suffer from all sorts of distribution biases, e.g., lacking some kinds of messages due to their rare uses in practice, leading to sub-optimal results. %In our case, NetPlier fails to recognize many OSDP packet fields.

For instance, NetPlier partitions the first four bytes of an OSDP packet
as | 0x53\ 0xff | 0x29 | 0x00 | ... |,
which mistakenly places the first two bytes into the same field
and splits $B[2]$ and $B[3]$ into two different fields while $B[2..3]$ (with the value of $B[3]B[2]$ that represents a two-byte integer with $B[3]$ the most significant byte and $B[2]$ the least.) should be a single field representing the packet length.
However,
since most input packets are shorter than 255, $B[3]$ is always zero while $B[2]$ has different values in different packets.
Thus, these two bytes follow different distributions in the packet samples, and NetPlier incorrectly regards them as separate fields.
Moreover, NetPlier does not infer semantic constraints such as the condition at Line 11 in Figure~\ref{fig:motivation-1}(a).
Such imprecise formats prevent a grammar-based protocol fuzzer from finding the zero-day (see \S\ref{sec:eval}) and fail to enhance Wireshark and Snort (see Appendix~\ref{app:wireshark}).

Category-two methods dynamically analyze protocol execution using a set of input packets. AutoFormat~\cite{lin2008automatic} is a representative. It leverages the observation that
most packet parsers utilize top-down parsing such that they invoke
a function to parse a sub-structure.
Therefore, the dynamic call graph in parsing a packet discloses its structure. However, the function call hierarchy may not
be sufficiently fine-grained to disclose detailed packet formats.
Similar to NetPlier, it does not infer semantic constraints across fields, such as the one at Line 11 in Figure~\ref{fig:motivation-1}(a).
As dynamic analysis,
the inferred format may be incomplete,
depending on the coverage of the input packets that drive the dynamic analysis.
For instance, in our evaluation,
Autoformat misses 15 out of the 27 packet types because these types of packets do not appear in regular workloads.
%As such, its results have only 74\% precision and 31\% recall (see \S\ref{sec:eval}).

Some category-two techniques, e.g., Tupni~\cite{cui2008tupni}, can precisely infer semantic constraints among packet fields.
However, 
as dynamic analyses,
they suffer from the innate coverage problem. As per our results, the inference results of Tupni may miss >50\% of possible formats.
The problem is that if the program executions analyzed by Tupni do not cover
the file-transfer command, i.e., Lines 5-15 in Figure~\ref{fig:motivation-1}(a),
Tupni will not generate formats for the command.
Without the formats, 
it is hard for a fuzzer to generate packets that can pass the validity check at Line 11 and expose the bug at Line~14.

\subsection{Our Solution and Security Applications}

Observing that the source code discloses substantial information about packet formats,
we propose a category-three method that lifts the source code of OSDP to the protocol formats.
For instance,
Line~5 of the code in Figure~\ref{fig:motivation-1}(a)
indicates that the command code for file transfer is 0x7c.
Line~7 indicates that $B[6]$ is a field representing the file type to transfer.
Lines~8-9 load two four-byte integers to variables \textsf{file\_size} and \textsf{file\_offset} and thus indicate that there are two four-byte fields, one from $B[7]$ to $B[10]$ and the other from $B[11]$ to $B[14]$,
meaning the size and the offset of the file to transfer, respectively.
In addition to the syntactic information (e.g., field partitioning),
the code also discloses the semantic relations across fields.
For instance, the if-statement at Line 11 implies a cross-field constraint
dictating that if $B[4]~\&~4 = 0$, a valid packet must satisfy the constraint $B[3]B[2] -7 \ge 12$ or, otherwise, $B[3]B[2] - 8 \ge 12$.

We extract the above syntactic and semantic information via static analysis and produce a BNF-style production rule in Figure~\ref{fig:motivation-1}(b).
%We resort to static analysis to address the coverage problem in dynamic analysis.
%As a result,
Our lifted formats are both precise and of high coverage.
In terms of precision,
we precisely identify each field and its name as shown in Figure~\ref{fig:motivation-1}(b)
and, meanwhile, also specify the field constraints as first-order-logic formulas.
In terms of coverage, since we do not rely on any input packets like category-one and category-two techniques,
any format included in the source code will be inferred.
%As reported in \S\ref{sec:eval}, the OSDP formats inferred
%by our tool have 100\% precision and recall.
We can use the lifted formats to support many applications.

\defpar{Application 1: Finding Zero-days by Network Fuzzing.}
We leverage a theorem prover such as Z3~\cite{de2008z3} to produce valuations for individual packet fields, such as the $B[i]$'s in Figure~\ref{fig:motivation-1}(b), which satisfy the semantic constraints.
The generated packets can pass the check at Line~11 in Figure~\ref{fig:motivation-1}(a),
thereby enabling the discovery of the CVE at Line~14.
In addition, our inferred packet format is of high coverage and allows the fuzzer to generate diverse packets to improve test coverage.
In particular, the vulnerable code can only be reached when the packet is a file-transfer command, i.e., the 0x7c branch of the switch statement (Line 5). If the format is not covered, 
the chance that a fuzzer can mutate a packet of different types to a valid file-transfer command is very slim. 
Our evaluation shows that the lifted formats can improve the coverage of fuzzing by 20-260\% and allow us to detect 41 more zero-days, compared to using the inferred specifications by category-one and category-two methods, which can only detect 12 zero-days.
Note that we do not claim direct contributions to fuzzing. Instead, our approach is orthogonal to existing fuzzing methods that rely on packet formats.

\defpar{Application 2: Network Traffic Auditing.}
Wireshark is the foremost protocol analyzer to ensure network security for hundreds of protocols~\cite{wireshark}.
Supporting a new protocol in Wireshark can be achieved by providing an extension, which is usually a library to parse protocol packets. We develop an extension generator that takes a lifted format as input and generates the corresponding Wireshark extension. 
Figure~\ref{fig:motivation-2}(c) shows that with the generated extension, Wireshark can look inside an OSDP packet sampled from the abnormal traffic.
Our lifted format provides not only precise packet syntax but also informative field names extracted from variable names.
With Wireshark, we observe that all packets during the abnormal traffic have the field \textsf{osdp.address}=35 and \textsf{osdp.cmd}=0x7c, indicating they are all from a device with the id \#35 via the file-transfer commands. 
As will be shown in our evaluation, 
category-two approaches miss over 50\% of possible fields. Extensions built from these incomplete formats would render Wireshark failing to process many received packets.
In addition,
they can hardly provide field names that are as informative as the ones we can provide.

\defpar{Application 3: Network Intrusion Detection.}
Due to the space limit, we put discussions of this application in Appendix~\ref{app:wireshark}.

\defpar{Remark.}
While our inferred formats have high precision and recall close to 100\%, 
like all previous works, there may still be missing or wrong formats, due to the inherent limitations of static program analysis (see \S\ref{subsec:limitations}). 
However, the inferred format still matters in practice, because many downstream applications do not require perfect formats.
For example, although format inaccuracies cause degraded efficacy improvement in fuzzing, the performance may still be far better than without any formats or having low-quality formats. This is similar for network traffic auditing and intrusion detection.

\startreviseblock
\section{Background and Overview}
\delete{\Large \textbf{3} \enskip\enskip \Large\textbf{LIFTING PROCEDURE OVERVIEW}}
\label{sec:scope}

\dismissreviseblock

\noindent
This section \revise{provides some background knowledge of our approach and} overviews the lifting procedure\delete{ and defines the inferred protocol formats}, in order to facilitate the later discussion with more complexity.

\begin{figure}[t]
    \centering
    \includegraphics[width=\columnwidth]{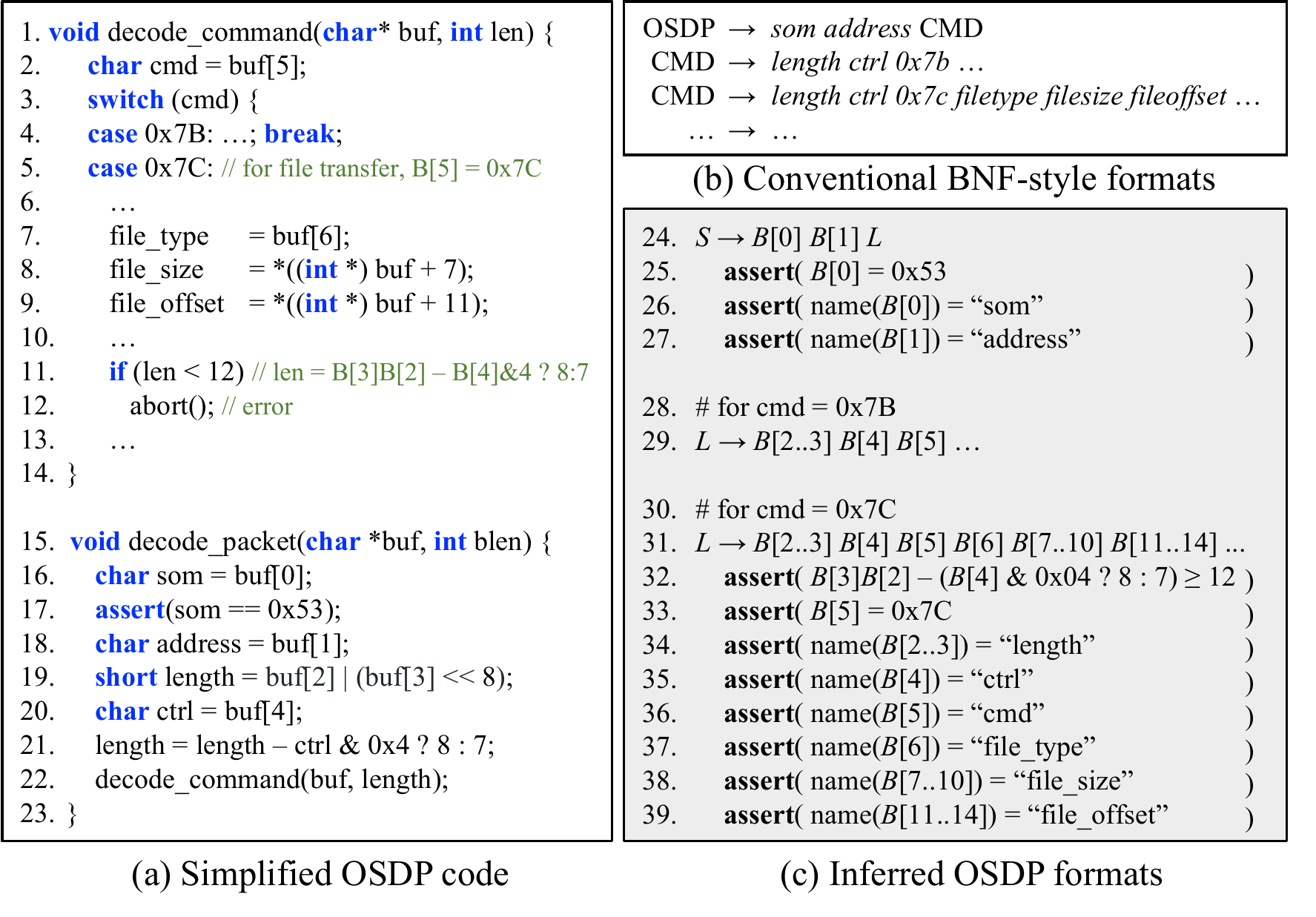}
    \caption{Extended example for OSDP.}
    \label{fig:osdp}
\end{figure}

\startreviseblock
\defpar{Protocol Format vs. Protocol Specification.}
Generally, the specification of a protocol consists of protocol formats and protocol state machines~\cite{narayan2015survey}.
Protocol formats are often specified using a grammar in BNF, which specifies how a network packet, i.e., a bit or byte stream,
can be dissected into multiple segments, i.e., fields, and specifies the semantic constraints the fields need to satisfy.
For example, Figure~\ref{fig:osdp}(b) shows a typical BNF-style format of OSDP packets.
The productions specify how an OSDP packet can be divided into multiple fields such as \textit{som}, \textit{address}, and so on.
Like many previous works~\cite{cui2007discoverer,kleber2018nemesys,kleber2020message,ye2021netplier,wang2011biprominer,wang2012semantics,luo2013position,gopinath2020mining,hoschele2016mining,caballero2009dispatcher,lin2008automatic,cui2008tupni,caballero2007polyglot,wang2009reformat,lin2010reverse,liu2013extracting},
\tool\ focuses on inferring the protocol formats.

Protocol state machines, on the other hand, specify the state transitions of a network server upon receiving or sending a specific network packet.
For instance, a TCP server may transition from the state SYN-SENT, meaning it waits for a matching connection, to the state ESTABLISHED, meaning a connection has been established, upon receiving an acknowledgment packet.
There have been many works also focusing on state machine inference~\cite{comparetti2009prospex,ye2021netplier,leita2005scriptgen,cui2006protocol,shevertalov2007reverse,antunes2011reverse,wang2011inferring,cho2010inference,zhang2012mining,laroche2012network,mcmahon2022closer,joeri2015protocol}.
Typically, these works accept a set of network packets as input and produce the protocol state machines.
While \tool\ only focuses on the formats,
since the formats are sufficient to produce a number of packets,
one can feed the packets into the aforementioned techniques for state machine inference.

\defpar{Static Program Analysis.}
Static analysis analyzes program behaviors without executing programs
and 
%is often referred to by 
often has various 
%terms 
forms such as dataflow analysis and abstract interpretation.
As pointed out by \citet{cousot1979systematic},
while dataflow analysis and abstract interpretation are in different forms,
%understand static analysis from different angles,
they are equivalent when used to compute sound results.
Basically, this is because both of them use abstract values to approximate program behavior.
A complete/sound analysis uses abstract values, which are often formulas over predefined symbols,
to under-/over-approximate concrete values that may assign to program~variables at runtime.
For instance, in our work, the abstract value of a variable is a formula over bytes, e.g., $B[0], B[1], \dots$, in a network packet.
Here, $B[i]$ is a byte ranging from 0x00 to 0xFF and, thus, over-approximates all possible values of the $(i+1)$th byte.

A static analysis is often defined by a set of transfer functions and merging functions over abstract values.
A transfer function specifies how we compute an abstract value when visiting a program statement.
For instance, given the abstract values of two variables, e.g., $x=B[1], y=B[2]$, the transfer function of the statement $z=x+y$,
accepts the abstract values of $x$ and $y$ as the input and outputs the abstract value of $z$, which typically is $B[1] + B[2]$.

When a variable is assigned multiple abstract values in two~program paths,
at the joint point of the two paths, we use a merging function to compute a merged abstract value.
For instance,
assume that we compute $x=B[1]$ and $x=B[2]$ in two different paths and $\Theta$ is an operator returns either of its operands.
At the joint point, the merged abstract value of $x$ could be $\Theta(B[1], B[2])$.
This value is sound as it over-approximates the value of $x$, saying that $x$ is either $B[1]$ or $B[2]$. However, it is not complete or not precise, because it loses the path information, i.e., from which path $x=B[1]$ (or~$B[2]$). 

A static analysis can be performed with varying degrees of precision.
In this work, we choose to perform a path-sensitive static analysis, which is of high precision as it can distinguish abstract values from different paths.
To this end, one often needs to enumerate all paths in a program, just like symbolic execution~\cite{king1976symbolic},
which, however, suffers from the notorious path-explosion problem due to the exponential number of paths in a program.
In this work,
we aim to significantly mitigate this problem by introducing a special merging operator, $\Theta_\kappa$, as explained later.

%Alternatively,
%to mitigate path explosion,
%one may introduce special operators in the aforementioned merging function so that multiple paths can be merged into one at the joint point.
%For instance,
%in our work, we use $\Theta_\kappa(B[1], B[2])$ to denote the merged abstract value of $x$ so that it keeps the path information, i.e., 
%$x=B[1]$ only when an if-statement uniquely identified by $\kappa$ takes the true branch and $x=B[2]$ otherwise.
%As such, if $x=\Theta_\kappa(B[1], B[2])$ and $z=\Theta_\kappa(B[3], B[4])$ at some program point,
%we can conclude that when $x=B[1]$, the value of $z$ must be $B[3]$, because they are from the same path that takes the true branch of the if-statement identified by $\kappa$.
%Without the identifier $\kappa$, we may mistakenly conclude that when $x=B[1]$, the value of $z$ may be $B[4]$.

\dismissreviseblock

\defpar{\revise{Input \& Output of \tool.}}
The input of our static analyzer is \revise{the source code of} a \textit{top-down protocol parser}~\cite{topdown} \revise{written in~C}.
\revise{A top-down parser applies each production rule in a BNF-style format to incoming bytes of the network packet, working from the left-most symbol of a production rule and then proceeding to the next production rule for each non-terminal symbol encountered~\cite{aho2007compilers}.}
Given \replace{the source code of the parser}{the parsing function of a protocol}, \revise{e.g., \textit{parse(char* buf, int len) \{ ... \}},}
the user annotates \revise{the parameters, i.e.,}
the \delete{packet }buffer variable\revise{, \textit{buf}, that contains the network packet to parse, and}
\replace{and the buffer-length}{the integer} variable\revise{, \textit{len}, which stands for the packet length}.
Except for the two annotations, \replace{our approach}{\tool} is fully automated.
\delete{In practice, a parsing function as well as the packet to parse can be easily found and annotated in two steps.
    First, we find the statements calling the standard APIs, e.g., recv(), that receive network packets.
    Second, once the received packet is found, we can easily identify the parsing function
    as the received packet is often immediately sent to the parsing function for further processing.}

The \mreplace{product}{output} of \mreplace{our static analysis}{\tool} is \mreplace{a BNF-style}{the protocol} format defined below.
The format is similar to common BNF so that it aligns well with existing standards in formally describing protocol formats.

\begin{definition} [Protocol Format]
    \label{def:spec}
    The format includes {syntax} and {semantics}.
    The syntax is denoted by production rules in BNF, where each rule is a sequence of consecutive bytes.
    Semantics is described by \revise{non-recursive} first-order-logic \revise{(FOL)} constraints \revise{with two special functions, \textsf{name(...)} and \textsf{repeat(...)}, which are explained in the example below}.
    The format satisfies three properties:
    \begin{enumerate}
        \item Each terminal symbol in the grammar is either $B[i]$ or~$B[i..j]$, which \replace{stands}{is a bit-vector standing} for the ($i+1$)th byte or a range of bytes from $B[i]$ to $B[j]$;
        \item Each production rule is associated with a set of assertions \revise{that assert FOL constraints} over the terminals in this rule. The \mreplace{assertions}{constraints} must not conflict with each other.
        \item Each assertion contains only a single \mreplace{literal, i.e., an atomic formula without}{atomic constraint that does not contain} any connectives $\land$ or $\lor$.
    \end{enumerate}
\end{definition}

\noindent\textbf{Example \revise{(Input of \tool)}.} 
Figure~\ref{fig:osdp} extends the example in Figure~\ref{fig:motivation-1}.
It shows a simplified OSDP parser starting from \mdelete{the entry function at }Line 15.
The packet is a byte array stored in \textit{buf} and the array length is \textit{blen}.
The user needs to annotate the two variables\mdelete{ for \tool}.
\revise{The parser with the annotations is the input of \tool.}
In Lines 16-21, the parser loads the first five bytes into the variables \textit{som}, \textit{address}, \textit{len}, and \textit{ctrl},
where \textit{som} stands for ``start of message'' and is used to identify OSDP packets.
The remaining code invokes the function
\textit{decode\_command} to parse an OSDP command \mrevise{as explained in Figure~\ref{fig:motivation-1}}.
\mdelete{This function consumes one byte to determine the command type in Line 3 and then parses the rest accordingly.}\hfill$\Box$

\smallskip
\noindent\textbf{\revise{Example (Output of \tool).}} 
Figure~\ref{fig:osdp}(b) shows a typical BNF-style format of OSDP, which is often manually constructed. 
\mdelete{The first production rule says that an OSDP packet consists of terminals {\it som} and {\it address}, which are followed by a non-terminal CMD. The next few rules define various types of CMD messages. For example, the file-transfer command is denoted by the third rule and  identified by the command code 0x7c.}
\mreplace{Our inferred}{The output} format \revise{of \tool} is shown in Figure~\ref{fig:osdp}(c), which closely resembles the manually constructed BNF in (b). 
The first rule in (c) resembles the first rule in (b), where we correctly determine that the first two bytes, $B[0]$ and $B[1]$, are two separate fields, corresponding to the fields \textit{som} and \textit{address} in (b).
Similarly,
the second and third rules in (c) resemble the two CMD rules in (b),
where, besides single-byte fields, we also correctly determine multi-byte fields including $B[2..3]$, $B[7..10]$,
and $B[11..14]$, corresponding to the fields \textit{length}, \textit{filesize},
and \textit{fileoffset} in (b).

\mreplace{Our inferred}{The output} format also associates each rule with two kinds of assertions.
One kind, such as Line 25 and Lines 32-33,
specifies the semantic constraints among packet fields.
They are inferred from branching conditions in the code.
When we infer a constraint including a value like $B[3]B[2]$,
it indicates a two-byte field with $B[3]$ the most significant byte and $B[2]$ the least.
In other words, 
in addition to \usmashit{field boundaries}, our format also expresses the \usmashit{endianness}, \mreplace{which}{whereas} the standard BNF cannot.
\tool{} also describes semantic constraints not expressible in standard BNF such as the one in Line 32.
All constraints have the \usmashit{bit-level precision}.
For instance,
the expression $B[4]~\&~4$ in Line 32 computes the third bit of $B[4]$.

The other kind, such as Lines 26-27 and Lines 34-39, specifies the field names,
which provide \usmashit{high-level field semantics} for us to understand the format.
In addition to those in the example,
we also produce many other names such as \fieldnamefun{$B[i..j]$} = \fieldname{timestamp}/\fieldname{checksum}
to indicate a timestamp/checksum field.
As explained later, we infer such high-level semantics using the names of program variables or library APIs. \hfill$\Box$

\smallskip
\mreplace{As illustrated above, our lifted format precisely matches the manually generated one and provides substantial additional information. 
We further}{In addition to the example above, we} elaborate on several places where our format is more expressive than the standard BNF.

\defpar{{Direction and Variable-Length Fields.}}
A direction field locates another field and
is often a length field,
whose value encodes the variable length of a target field~\cite{caballero2007polyglot}.
\mdelete{A variable-length field in our format is denoted by a byte sequence, $B[i..i+n]$,
where $i$ and $n$ may not be constants.}For instance, we may produce a rule
$S\rightarrow B[0]B[1..B[0]]$,
where $B[1..B[0]]$ is a variable-length field whose length is determined by the direction field $B[0]$. 

\defpar{{Repetitive Fields.}}
\mreplace{We use assertions to describe}{Our format can also specify} repetitive fields and \mdelete{specify }how many times a field repeats.
For instance, 
we may produce the production rule $S \rightarrow B[0]B[1..2]B[3]$ with three assertions: (1) $\textbf{assert}(B[0] = B[3] = 0x00)$,
(2) $\textbf{assert}(B[1]B[2] \ne 0x0000)$,
and (3) $\textbf{assert}(\textsf{repeat}(B[1..2]) = 3)$.
The first assertion constrains the first and last byte.
The second constrains the field $B[1..2]$ in middle.
The third states that the middle field repeats three times.
When generating packets based on the rule,
we first generate a packet satisfying the first two assertions,
e.g., $0x00~0x0001~0x00$.
Due to the third assertion,
we insert another two fields satisfying the same constraints as $B[1..2]$,
e.g., $0x00~0x0001~\textcolor{red}{0x0011}~\textcolor{blue}{0x0101}~0x00$.

%% file: overview.tex
\section{Technical Challenges} 
\label{sec:nutshell}

\begin{figure*}[t]
    \centering
    \includegraphics[width=\textwidth]{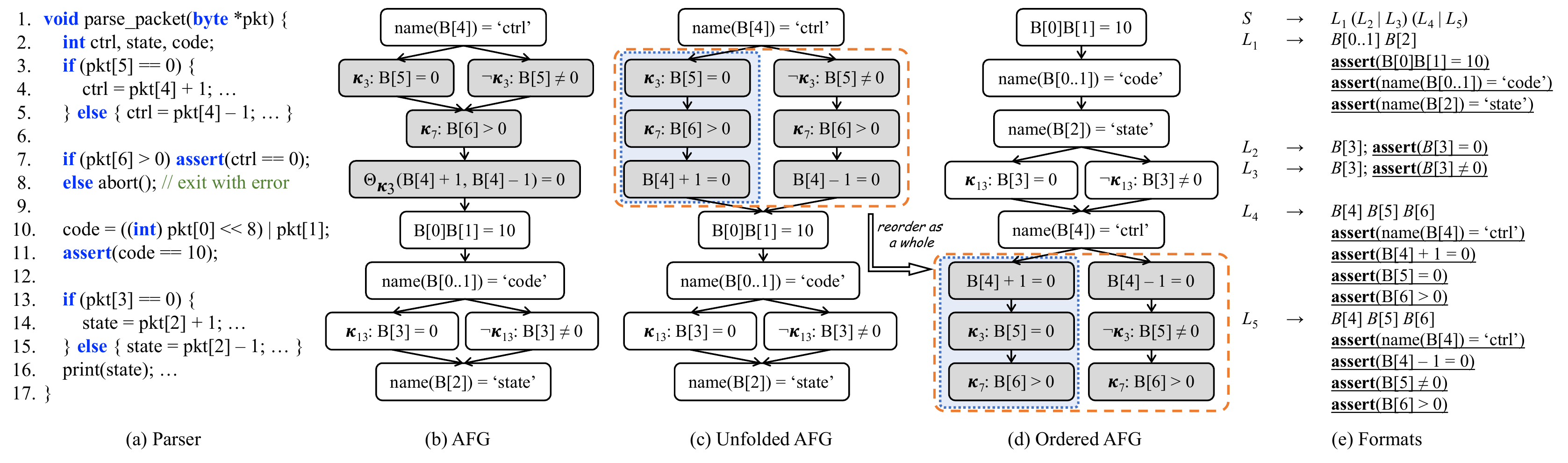}
    \caption{A crafted running example.}
    \label{fig:overview}
    \vspace{-12pt}
\end{figure*}

This section discusses two prominent challenges as well as our ideas for addressing them. It provides a context for the detailed discussion later and is driven by the crafted running example in Figure~\ref{fig:overview}. Figure~\ref{fig:overview}(a) shows a protocol parser and Figure~\ref{fig:overview}(e) shows an ideal lifted format. The code implies complex cross-field constraints which are clearly represented in Figure~\ref{fig:overview}(e). We can see that a packet of the protocol contains three segments $L_1$, $L_2 | L_3$, and $L_4 | L_5$. The segment $L_1$ has two fields, a {\it code} field  (reflected by Lines 10-11 in the code) and a {\it state} field (reflected by Lines 14-16 in the code). 
Both $L_2$ and $L_3$ consist of a single field located at byte offset 3, and differ only in the field value (reflected by Line 13 in the code).
Both $L_4$ and $L_5$ consist of three single-byte fields and differ in the semantic constraints (reflected by Lines 3-8 in the code).

\defpar{Challenge 1: Insufficiency of Traditional Static Analysis.}
Traditional static analysis is path-insensitive and merges analysis results from different paths at their joint point to achieve scalability. \mreplace{It is well-known that}{As introduced before,} such merging yields over-approximation and incurs low precision.
For example, the abstract values of {\it ctrl} from the two branches at Lines 4 and 5, respectively, are merged at Line~6, yielding $\textit{ctrl}= B[4]+1 \vee \textit{ctrl}= B[4]-1$. As such, we lose the correlation between $B[4]$ and $B[5]$ as the precise value of \textit{ctrl} should depend on the value of $B[5]$ due to the if-statement at Line 3.
\revise{In consequence, the resulting format will lose the correlations between $B[4]$ and $B[5]$, while in the ideal format shown in Figure~\ref{fig:overview}(e), the production rules $L_4$ and $L_5$ include such correlation, i.e., $B[4] + 1 = 0 \Leftrightarrow B[5] = 0$ and $B[4] - 1 = 0 \Leftrightarrow B[5] \ne 0$.}

\delete{As a result, we may produce two production rules without constraining $B[5]$:}

\delete{$L_{4}' \rightarrow~ B[4]\ B[5]\ B[6]  \textbf{assert}(B[6]>0);~\textbf{assert}(B[4]+1=0)$}

\delete{$L_{5}' \rightarrow~ B[4]\ B[5]\ B[6] \textbf{assert}(B[6]>0);~\textbf{assert}(B[4]-1=0)$}

\delete{Compared to the ideal format in Figure~\ref{fig:overview}(e), especially $L_4$ and $L_5$, $L_4'$ and $L_5'$ are over-approximate, allowing invalid packets such as those satisfying $B[5]=0\wedge B[4]-1=0$.}

A typical solution is to use a {path-sensitive} static analysis that separately analyzes individual paths and does not merge results from multiple branches.
Lifting is thus reduced to enumerating paths, each constituting a production rule.
In our example, there are four paths that denote valid packets, i.e., 
{(P1)} $... \rightarrow4\rightarrow ... \rightarrow14\rightarrow...$, 
{(P2)} $... \rightarrow5\rightarrow... \rightarrow14\rightarrow...$,
{(P3)} $... \rightarrow4\rightarrow...  \rightarrow15\rightarrow...$, and 
{(P4)} $... \rightarrow5\rightarrow... \rightarrow15\rightarrow...$.
Thus, the lifted format has four rules, each of which corresponds to a path constraint.
For example, the format for the path {P1} is shown below.

\ffbox{\small
    \begin{minipage}{0.96\columnwidth}\centering
        \vspace{0.5mm}
\hspace{-20mm}$S \rightarrow  B[0..1]\ B[2]\ B[3]\ B[4]\ B[5]\ B[6]$\vspace{1mm}
\\ \hspace{10mm}$\textbf{assert}({\small B[5] = 0});  \textbf{assert}({\small B[6] > 0}); \textbf{assert}({\small B[4] + 1 = 0});$\vspace{1mm}
\\ \hspace{-8mm}$\textbf{assert}({\small B[0]B[1] = 10}); \textbf{assert}({\small B[3] = 0});$
\vspace{0.5mm}
    \end{minipage}
}

Enumerating program paths incurs the notorious path-explosion problem,
which has two consequences: (1) the analysis is not scalable and (2) the lifted format has an explosive number of semantic constraints.
For example, due to path explosion,
KLEE~\cite{cadar2008klee}, a state-of-the-art path-sensitive static analyzer, cannot finish analyzing Linux's implementation of L2CAP, a Bluetooth protocol containing a few thousand lines of code, within twelve hours. 
%Furthermore, compared to the ideal specification in Figure~\ref{fig:overview}(e), the above format is difficult to interpret for humans.

\defpar{Solution 1: Localized Path-Sensitive Analysis.}
We observe that in lifting, path-sensitivity is only needed in certain places. In our example, we only want to analyze sub-paths
$3\rightarrow 4\rightarrow 7$ and $3\rightarrow 5\rightarrow 7$ separately such that we can generate production rules $L_4$ and $L_5$ in Figure~\ref{fig:overview}(e).
%However, we do not want full path-sensitivity. 
{\em The criterion to determine a local code region for path-sensitive analysis is that the path conditions within the region and their branches have inter-dependencies.} For example, the condition at Line 3 determines the value of {\it ctrl}, which is checked inside the true branch at Line 7, allowing Lines 3-8 to form a region for path-sensitive analysis. In contrast, Lines 10-16 have no dependencies on Lines 3-8 and, thus, are considered separately.  
%\footnote{Strictly speaking, there is control dependence due to the abort and assert functions. However, such dependence is suppressed during our lifting as it merely denotes invalid packet instead of true dependence.} 
In Section~\ref{sec:approach}, we will discuss how we use a new selection operator and a novel representation called abstract format graph (AFG) to identify the regions for localized path-sensitive analysis.

\defpar{Challenge 2: Handling Out-of-Order Fields.}
Protocol parsers may not parse network packets in strict byte order.
Hence, if a naive lifting algorithm directly derives format from code, for example, generating production rules following the order that the bytes are accessed along program paths, the resulting format may have out-of-order fields, which do not comply with the BNF standard.  
For example in Figure~\ref{fig:overview}(a), the parser parses bytes 
$B[4..6]$ before $B[0..3]$.
Therefore, we have to break the program order. This requires us to reorder the bytes such that they follow the byte order while not violating program semantics.
For example, in Lines 3-8 in Figure~\ref{fig:overview} (a),
the access of {\it pkt[4]} occurs after that of {\it pkt[5]}. One cannot simply relocate Line 4 and the else branch in Line 5 to in front of Line 3, because the resulting program is broken as shown in the following.

\ffbox{
    \begin{minipage}{0.964\columnwidth}\centering
        \noindent
        \textsf{ctrl=pkt[4]-1;\enskip ctrl=pkt[4]-1;\enskip \textbf{if} (pkt[5]==0)\enskip ...}
    \end{minipage}
}

\defpar{Solution 2: Graph-Based Reordering.}
We propose to first abstract the code to the aforementioned AFG that models only the packet format-related behaviors and precludes the rest. As such, we do not need to transform the program which is complex and unnecessary. An algorithm is developed to ensure dependencies can be respected during reordering.

%% file: approach.tex
\section{Design}
\label{sec:approach}

Figure~\ref{fig:overview}(a-e) presents the workflow of \tool. 
Given the code of a protocol,
abstract interpretation is performed to construct 
an abstract format graph (AFG). Path-sensitive analysis is performed in selected local regions of AFG to produce an unfolded AFG, which is further reordered 
and post-processed to produce the lifted formats.

\subsection{Abstract Format Graph}\label{subsec:afg}

AFG is a directed acyclic graph representing first-order-logic constraints.
The AFG of a constraint $\rho$ is inductively defined by $\textsf{AFG}(\rho)$:
\begin{enumerate}
    \item $\textsf{AFG}(\textit{atomic constraint}) = \textsf{Vertex}(\textit{atomic constraint})$;
    \item $\textsf{AFG}(\rho_1 \land \rho_2) = \textsf{AFG}(\rho_1) \bowtie \textsf{AFG}(\rho_2)$;
    \item $\textsf{AFG}(\rho_1 \lor \rho_2) = \textsf{AFG}(\rho_1) \uplus \textsf{AFG}(\rho_2)$.
\end{enumerate}
\revise{Generally, a vertex of AFG is an atomic constraint that does not contain any connectives $\land$ or $\lor$,
and an edge means logical conjunction. In the definition, t}\delete{T}he first rule returns a single vertex \replace{containing the constraint $\rho$ if $\rho$ is a literal, i.e., without any connectives $\land$ or $\lor$}{for any atomic constraint}.
The second creates a graph for conjunction by connecting all \usmashit{exit vertices} (vertices without outgoing edges) of $\textsf{AFG}(\rho_1)$ to all \usmashit{entry vertices} (vertices without incoming edges) of $\textsf{AFG}(\rho_2)$.
The third creates a graph for disjunction by simply creating a union of the two graphs,
which contains the vertices and edges from both.\mdelete{ An AFG can be constructed to denote all path constraints in a program, and each AFG path \usmashit{represents} the path constraint of an individual program path,
i.e., conjoining the constraints in an AFG path yields an individual
path constraint.}
The following lemma states \mreplace{this}{the equivalence} relation between the graph
$\textsf{AFG}(\rho)$ and the constraint $\rho$.
\mrevise{In other words,
$\textsf{AFG}(\rho)$
 is an equivalent graphic representation of the constraint $\rho$.}
We put the proofs of all our lemmas in Appendix~\ref{app:proof}.

\newcommand{\LemmaAFG}{
\mdelete{Given the path constraint $\rho_i$ of each individual program path
    and any constraint $\rho$ equivalent to $\bigvee_i \rho_i$,
    each path in \textsf{AFG}($\rho$) represents a constraint $\delta$
    such that $\exists\rho_i.\delta=\rho_i$ or $\delta=\textit{false}$
    and, for any $\rho_i\ne\textit{false}$,
    there is an AFG path representing it.}\mrevise{Given $\textsf{AFG}(\rho)$ with $n$ paths,
we have $\rho \equiv\bigvee_{i=1}^n\rho_i$~where each $\rho_i$ equals the conjunction of all constraints in an AFG path.}\mremark{This revision simplifies the original lemma into an equivalent one}}
\begin{lemma}
    \LemmaAFG
    \label{lemma:afg}
\end{lemma}

\noindent
{\bf Example.}
\mrevise{Consider the constraint $\rho\equiv(a\lor b)\land c \land (d\lor e)$.
By definition, $\textsf{AFG}(\rho)$ is a directed graph with five nodes,
which respectively correspond to the five atomic constraints $a$, $b$, $c$, $d$, and $e$.
The AFG also contains four edges respectively from $a$ to $c$, from $b$ to $c$,
from $c$ to $d$, and from $c$ to $e$.
The AFG has four paths,
respectively representing four constraints,
$\rho_1 = a\land c \land d$, $\rho_2 = a\land c \land e$,
$\rho_3 = b\land c \land d$, and $\rho_4 = b\land c \land e$.
Apparently, we have $\rho_1\lor \rho_2 \lor \rho_3 \lor \rho_4 \equiv\rho$.
Thus, we say the $\textsf{AFG}(\rho)$ is an equivalent graphic representation of the constraint $\rho$.}\mdelete{Consider the code in Figure~\ref{fig:overview}(a) and the AFG in (b),
where $\Theta_{\kappa}$ is an operator we will introduce later
but, briefly speaking, $\Theta_{\kappa_i}(v_1, v_2)$
returns $v_1$ if if-statement at Line $i$ takes the true branch and returns $v_2$ otherwise.
Each AFG path like the one going through $B[5] = 0$ and $B[3] = 0$
represents the path condition of the path
$3\rightarrow4\rightarrow7\rightarrow13\rightarrow14$ in the code.}\mremark{We replace it with a new example because the old example uses the symbol $\Theta_\kappa$ which has not been formally defined.}
\hfill $\Box$

% \begin{figure}[t]
%     \centering
%     \includegraphics[width=0.95\columnwidth]{figs/workflow}
%     \vspace{-2mm}
%     \caption{Workflow.}
%     \vspace{-4mm}
%     \label{fig:workflow}
% \end{figure}

\subsection{Abstract Interpretation}\label{subsec:ai}
The static analysis derives an AFG denoting path constraints related to the packet format. 
It features a new selection operator at the joint point of branches, which enables localized path-sensitive analysis.

\defpar{Abstract Language.}
For clarity, we use a C-like language in Figure~\ref{fig:design_lang} to model our target programs.
%We discuss how we handle loops at the end of this section.
A program in the language has an entry function that parses an input network packet, \textit{pkt},
which is a byte array.
The parsing function often has a parameter specifying the packet length, \textit{len},
to avoid out-of-bounds access during parsing.
The language contains assignments, binary operations, statements that read bytes from the packet, assertions,
branching, and sequencing.
Each branching statement is labeled by a unique~identifier $\kappa$.
Although we do not include function calls or returns for discussion simplicity, 
our system is inter-procedural as a call statement is equivalent to a list of assignments from the actual parameters
to the formals, and a return statement is an assignment from the return value to its receiver.
\mrevise{The language includes statements reading bytes from the packet
but does not include statements that store values into the packet. This is because, for parsing purposes, the input packet is often read-only.
Note that the abstract language serves for demonstrating how we address the challenges discussed in \S\ref{sec:nutshell}.
Thus, for simplicity, we abstract away some common program structures, e.g., pointers and loops, from the language.
Dealing with these structures is not our technical contribution.
In \S\ref{subsec:impl}, we discuss how we handle them in our implementation.}\mdelete{Except for reading bytes from the packet, e.g., $v_1 \leftarrow \textit{pkt}[v_2]$,
the language abstracts away all other
pointer operations because the pointer analysis is not our
technical contribution and, in the implementation, we follow
existing works to resolve pointer relations.%~\cite{babic2008calysto}.
Note that, for parsing purposes, the input network packet, \textit{pkt}, is often read-only.
Thus, it does not have statements that store values into the~packet.}

\defpar{Abstract Domain.}
An abstract value of a variable represents all possible concrete values that may be assigned to the variable during program execution.
The abstract domain specifies the limited forms of an abstract value.
In our analysis,
the abstract value of a variable $v$ is denoted as $\tilde{v}$
and defined in Figure~\ref{fig:design_absval}.
An abstract value could be a constant or a special value \textit{length} that represents the packet length.
The $(\tilde{v} + 1)$th byte of the input packet is $B[\tilde{v}]$.
We introduce a new selection operator $\Theta_{\kappa}$ such that $v = \Theta_{\kappa}(v_1, v_2)$, which
means that when the if-statement at $\kappa$ takes the true branch, we have $v=v_1$,
 $v=v_2$ otherwise.
One may find that the operator $\Theta_{\kappa}$ is similar to the operator $\phi$ in the classic SSA code form~\cite{cytron1989efficient}
because both of them merge values from multiple branches. We note that
$\Theta_{\kappa}$ differs from $\phi$ in two aspects. First, in the SSA form, $v=\phi(v_1, v_2)$ is always placed at the end of a branching statement, whereas in our analysis $v=\Theta_{\kappa_i}(v_1, v_2)$ represents an abstract value of the variable $v$ and is propagated to many other places where the variable $v$ is referenced.
%, e.g., in the true branch at Line 7. 
Second, since $v=\Theta_{\kappa}(v_1, v_2)$ may be used at any place in the code,
we use the subscript $\kappa$ to record the branching statement where it originates. This is a critical design for the next step, i.e., the localized graph unfolding, as illustrated later.
%The inference rules (1)-(5) in Figure~\ref{fig:design_absval}
%can be easily understood with the semantics of $\Theta_{\kappa}$ explained. 
%As discussed in \S\ref{sec:nutshell},
%we support $\Theta_\kappa$-merged values,
%each of which
%returns its first operand if the $\textbf{if}_\kappa$-statement takes the true branch
%or, otherwise, the second.
%The key role of the operator $\Theta_\kappa$ in our design and its differences from the operator $\phi$ in classic SSA form  have been discussed in \S\ref{sec:nutshell}. Thus, we do not repeat it here.
An abstract value can also be a first-order logic formula over other abstract values.
To ease the explanation,
we only support binary formulas.

\begin{figure}[t]
    \centering
    \footnotesize
    \begin{tabular}{rcll}
        \textit{Function}~$F$ &:=&$\textit{parse}(\textit{pkt}, \textit{len}) \{~S;~\}$ &~\\
        \textit{Statement}~$S$ &:=& $v_1 \leftarrow v_2$ & \textbf{~::assign}\\
        & ~ & $~|~v_1 \leftarrow v_2 \oplus v_3$& \textbf{~::binary}\\
        & ~ & $~|~v_1 \leftarrow \textit{pkt}[v_2]$& \textbf{~::read}\\
        & ~ & $~|~\textbf{assert}(v_1)$ & \textbf{~::assertion}\\
        & ~ & $~|~\textbf{if}_\kappa~(v)~\{ S_1; \}~\textbf{else}~\{ S_2; \}$& \textbf{~::branching}\\
        & ~ & $~|~S_1;S_2$ & \textbf{~::sequencing}\\
        &&&\\
        & \multicolumn{3}{l}{$\oplus\in \{ \land, \lor, +, -, >, <, =, \ne, \dots \}$} %\circ, \triangledown, 
    \end{tabular}
    \caption{Language of target programs.}
    \vspace{-2mm}
    \label{fig:design_lang}
\end{figure}

\begin{figure}\footnotesize\centering
    \begin{tabular}{rcll}
        \textit{Abstract Value}~$\tilde{v}$ &:=& $c$ & \textbf{~::constant}\\
        %        & ~ & $~|~c$ & \textbf{~::constant}\\
        & ~ & $~|~\textit{length}$ & \textbf{~::packet length}\\
        & ~ & $~|~B[\tilde{v}]$ & \textbf{~::byte in packet}\\
        %        & ~ & $~|~m_\textbf{mid}$ & \textbf{~::base address}\\
        & ~ & $~|~\Theta_\kappa(\tilde{v}_1, \tilde{v}_2)$& \textbf{~::selection}\\
        & ~ & $~|~\tilde{v}_1 \oplus \tilde{v}_2$& \textbf{~::binary operation}
    \end{tabular}\\~\\
    $$
    \inference{
        \tilde{v}_1 = \Theta_\kappa(\tilde{v}_2,  \tilde{v}_2)
    }
    {
        \tilde{v}_1  = \tilde{v}_2
    }[(1)]
    \quad
    \inference{
        \tilde{v}_1 = \Theta_\kappa(\tilde{v}_2,  \tilde{v}_3)
    }
    {
        \tilde{v}_1\oplus \tilde{v}_4 = \Theta_\kappa(\tilde{v}_2 \oplus \tilde{v}_4, \tilde{v}_3 \oplus \tilde{v}_4)
    }[(2)]
    $$
    \\
    
    $$
    \inference{
        \tilde{v}_1 = \Theta_\kappa(\tilde{v}_2,  \tilde{v}_3)
    }
    {
        B[\tilde{v}_1] = \Theta_\kappa(B[\tilde{v}_2], B[\tilde{v}_3])
    }[(3)]
    $$
    \\
    $$
    \inference{
        \tilde{v}_1 = \Theta_\kappa(\Theta_\kappa(\tilde{v}_2, \tilde{v}_3),  \tilde{v}_4)
    }
    {
        \tilde{v}_1 = \Theta_\kappa(\tilde{v}_2,  \tilde{v}_4)
    }[(4)]
    \quad
    \inference{
        \tilde{v}_1 = \Theta_\kappa(\tilde{v}_2,  \Theta_\kappa(\tilde{v}_3, \tilde{v}_4))
    }
    {
        \tilde{v}_1 = \Theta_\kappa(\tilde{v}_2,  \tilde{v}_4)
    }[(5)]
    $$
    \caption{Abstract values.}
    \label{fig:design_absval}
\end{figure}

Figure~\ref{fig:design_absval} lists
the rules that normalize expressions over abstract values.
Rule~(1) states that we do not need a $\Theta_\kappa$ operator if 
we merge two equivalent values.
Rules~(2-3) state that any operation with a $\Theta_\kappa$-merged value is equivalent to operating on each value merged by the $\Theta_\kappa$ operator.
Rules~(4-5) simplify nested $\Theta_\kappa$ operators.

\begin{figure*}[t]
    \centering
    \begin{minipage}[t]{\textwidth}\footnotesize
$$
\inference{
   \mathbb{E} = \emptyset \enskip\enskip\enskip \mathbb{G} = \emptyset
}
{
    \mathbb{E},\mathbb{G}\vdash \textit{parse}(\textit{pkt}, \textit{len}): \mathbb{E}[\textit{len}\mapsto\textit{length}], \mathbb{G}
}[\textbf{init}]
\quad\quad\quad
\inference{
   \mathbb{E}(v_2) = \tilde{v}_2
}
{
   \mathbb{E}, \mathbb{G} \vdash v_1 \leftarrow v_2: \mathbb{E}[v_1\mapsto \tilde{v}_2], \mathbb{G}
}[\textbf{assign}]
\quad\quad\quad
\inference{
    \mathbb{E}(v_2) = \tilde{v}_2\enskip\enskip\enskip \mathbb{E}(v_3) = \tilde{v}_3
}
{
    \mathbb{E}, \mathbb{G} \vdash v_1 \leftarrow v_2\oplus v_3: \mathbb{E}[v_1\mapsto \tilde{v}_2 \oplus \tilde{v}_3], \mathbb{G}
}[\textbf{binary}]
$$
\\
$$
\inference{
   \mathbb{E}(v_2) = \tilde{v}_2
}
{
    \mathbb{E}, \mathbb{G} \vdash v_1 \leftarrow \textit{pkt}[v_2]: \mathbb{E}[v_1\mapsto B[\tilde{v}_2]], \mathbb{G}
}[\textbf{read}]
\quad\quad\quad
\inference{
    \mathbb{E}(v_1) = \tilde{v}_1
}
{
    \mathbb{E}, \mathbb{G} \vdash \textbf{assert}(v_1): \mathbb{E}, \mathbb{G}\bowtie \textsf{AFG}(\tilde{v}_1)
}[\textbf{assertion}]
\quad\quad
\inference{
    \mathbb{E}_1, \mathbb{G}_1 \vdash S_1: \mathbb{E}_2, \mathbb{G}_2
    \enskip\enskip\enskip
    \mathbb{E}_2, \mathbb{G}_2 \vdash S_2: \mathbb{E}_3, \mathbb{G}_3
}
{
    \mathbb{E}_1, \mathbb{G}_1 \vdash S_1;S_2: \mathbb{E}_3, \mathbb{G}_3
}[\textbf{sequencing}]
$$
\\
$$
        \inference{
            \mathbb{E}(v) = \tilde{v} 
            \enskip\enskip\enskip 
            \mathbb{G}_1=  \textsf{AFG}(\tilde{v})
            \enskip\enskip\enskip
            \mathbb{G}_2= \textsf{AFG}(\lnot \tilde{v})
            %\enskip\enskip\enskip
            \\
            \mathbb{E}, \mathbb{G} \bowtie \mathbb{G}_1
            \vdash S_1: 
            \mathbb{E}_{\kappa}, \mathbb{G} \bowtie \mathbb{G}_{\kappa}
            \enskip\enskip \enskip
            \mathbb{E}, \mathbb{G} \bowtie \mathbb{G}_2, 
            \vdash S_2:  
            \mathbb{E}_{\lnot\kappa}, \mathbb{G} \bowtie \mathbb{G}_{\lnot\kappa}
        }
        {
            \mathbb{E}, 
            \mathbb{G}
            \vdash 
            \textbf{if}_\kappa~(v)~\{ S_1; \}~\textbf{else}~\{ S_2; \}: 
            \textsf{mergeE}(\mathbb{E}_{\kappa},\mathbb{E}_{\lnot\kappa}, \kappa), 
            \mathbb{G} \bowtie (\mathbb{G}_{\kappa}\uplus \mathbb{G}_{\lnot\kappa})
        }[\textbf{branching}]
    \quad
        \fbox{\parbox{7.6cm}{
                $\mbox{ }$
                \textbf{procedure} \textsf{mergeE}($\mathbb{E}_1$, $\mathbb{E}_2$, $\kappa$)\\
                $\mbox{    }\mbox{    }\mbox{    }\mbox{    }\mbox{    }\mbox{    }\mbox{    }\mbox{    }$\textbf{foreach} $(v, \tilde{v}_1)\in \mathbb{E}_1 \land (v, \tilde{v}_2)\in \mathbb{E}_2$ \textbf{do   } $\mathbb{E}_1\leftarrow\mathbb{E}_1 [v\mapsto \Theta_\kappa(\tilde{v}_1, \tilde{v}_2) ]$;
                \\
                $\mbox{    }\mbox{    }\mbox{    }\mbox{    }\mbox{    }\mbox{    }\mbox{    }\mbox{    }$\textbf{return} $\mathbb{E}_1$;
        }}
$$
	\end{minipage}
    \vspace{1mm}
    \caption{Inference rules and auxiliary procedure.}
    \vspace{-1mm}
    \label{fig:semantics_basic}
\end{figure*}

\defpar{Abstract Semantics.}
The abstract semantics describe how we analyze a given protocol parser.
They are described as transfer functions of program statements.
Each transfer function updates the program's abstract state,
which is a pair
$(\mathbb{E}, \mathbb{G})$.
Given the set $V$ of program variables and the set $\tilde{V}$ of abstract values,
$\mathbb{E}: V\mapsto \tilde{V}$ maps a variable to its abstract value.
We use $\mathbb{E}[v\mapsto \tilde{v}]$ to denote updating the abstract value of the variable $v$ to $\tilde{v}$.
$\mathbb{G}$ is the output AFG.
Since AFG is an equivalent form of path constraint,
we directly create AFG without computing the path constraint first.

Figure~\ref{fig:semantics_basic} lists the transfer functions as inference rules.
In each rule, the part above the horizontal line includes a set of assumptions
and, under these assumptions, the bottom part describes the abstract states before and after a statement $S$,
in the form of
$\mathbb{E}, \mathbb{G} \vdash S: \mathbb{E}', \mathbb{G}'$.
Initially,
we assign the special abstract value
$\textit{length}$ to the variable \textit{len}, which represents the length of input network packet.
The rules for assignment, binary operation, read operation, and assertion are straightforward.
For instance, in the rule for assertions,
the abstract value $\tilde{v}_1$ represents a constraint that must be satisfied.
Therefore, we append the graph $\textsf{AFG}(\tilde{v}_1)$ to the graph $\mathbb{G}$.
This is equivalent to appending the constraint $\tilde{v}_1$ to the current path constraint.
%Observe that nodes and edges are only admitted to the resulting AFG through this rule. 
%That is, if an abstract value is not directly/transitively used in a validity check assertion, it is precluded from the resulting AFG. This allows handling the challenge 2 in \S\ref{sec:nutshell}. 
%\xz{fix and check, this should address your concern about challenge 2}\qs{not true as discussed last Friday}

The sequencing rule states that, for two consecutive statements,
we analyze them in order, using the postcondition of the first statement as the precondition for the second.
In the branching rule,
$\mathbb{G}$ denotes the path constraint before the branching statement.
$\mathbb{G}_1$ and $\mathbb{G}_2$ represent the branching condition and its negation.
Thus,
$\mathbb{G}\bowtie\mathbb{G}_1$ and $\mathbb{G}\bowtie\mathbb{G}_2$
represent the initial path constraints before the two branches.
After analyzing the two branches,
the resulting AFGs are assumed to be $\mathbb{G}\bowtie\mathbb{G}_{\kappa}$ and $\mathbb{G}\bowtie\mathbb{G}_{\lnot\kappa}$.
The branching rule states that, under these assumptions and after an $\textbf{if}_\kappa$-statement,
we merge the abstract states from both branches.
The procedure \textsf{mergeE} merges abstract values of the same variable via the $\Theta_\kappa$ operator.
Graph merging is straightforward based on the definition of AFG,
which is equivalent to
merging path constraints of the two branches with the common prefix pulled out. Our merging is different from the value merging in traditional analyses due to the use of the selection operator. On one hand, merging allows achieving scalability as the number of values is no longer exponential of the number of statements. On the other hand, the selectors in abstract values can be unfolded to support path-sensitive analysis if needed.

\vspace{1.5mm}
\noindent\textbf{Packet Fields.}
The abstract interpretation builds the AFG to represent the path constraints.
As discussed in \S\ref{sec:scope},
from these constraints,
it is direct to infer the endianness, field boundaries, and direction fields.
For instance, if multiple consecutive bytes, e.g., $B[0]$ and $B[1]$ in Figure~\ref{fig:overview}, belong to a single field, 
the field value, e.g., $B[0]B[1]$, will be computed and occur in the path constraint. 

\vspace{1.5mm}
\noindent\textbf{High-Level Field Semantics.}
We also extend our analysis to infer high-level field semantics, i.e., field names, using rich source code information.
Such high-level semantics can help better understand a format, e.g., identifying checksum fields and distinguishing keywords and delimiters (both of which are constant fields).
As illustrated in Figure~\ref{fig:overview}, we can name a field (via some variable name) by adding extra path constraints.
Formally, given the AFG $\mathbb{G}$ and a formula over a field $B[i..j]$, denoted as $\textsf{f}(B[i..j])$,
we name the field by $\mathbb{G}\bowtie \textsf{AFG}(\fieldnamefun{B[i..j]}=\fieldname{var})$
if there is a statement assigning $\textsf{f}(B[i..j])$ to the variable $\textsf{var}$.
In addition to variable names, we also leverage system APIs used in the code.
For instance,
if a field $B[i..j]$ is used in the system API, \textsf{difftime()},
it is likely to be a timestamp field.
In our experience,
this method helps us identify many special fields via names
such as  `length', `version', `checksum', `timestamp', etc.
In our current implementation, we handle all standard C APIs.
If there are multiple options for naming a field, we prefer the names inferred by system APIs
because software developers may not be careful to name program variables.
If there are still multiple options, we simply keep the first. 

\defpar{Example.}
Given the code in Figure~\ref{fig:overview}(a), the abstract interpretation yields the AFG in (b) from top to bottom.
After Line 5, we merge the two paths forked from Line 3 and get the path constraint: 
$\rho\equiv \fieldnamefun{B[4]} = \fieldname{ctrl} \land (B[5] = 0\lor B[5] \ne 0).$
By the branching rule,
we do not compute the path constraint but directly create the equivalent AFG,
i.e., the first two rows in Figure~\ref{fig:overview}(b).
We name the byte $B[4]$ \fieldname{ctrl} because the arithmetic results of $B[4]$ are assigned to the variable \varname{ctrl} in both branches.
The constraint $B[5] = 0\lor B[5] \ne 0$ merges the branching constraints.
%It is not simplified to \textit{true} because, if simplified, we will miss two possible packet formats --- the protocol may distinguish two different packet formats based on whether the value of $B[5]$ is zero.
Meanwhile,
the abstract store is updated such that 
$\varname{ctrl} = \Theta_{\kappa_3}(B[4] + 1,  B[4] - 1)$.
% ,
% where $v = \Theta_{\kappa_3}(v_1, v_2)$
% means that when the if-statement at Line~3 takes the true branch, we have $v=v_1$,
%  $v=v_2$ otherwise.
At Line 7, since the false branch aborts, we only consider the true branch,
for which we add $B[6] > 0 \land \Theta_{\kappa_{3}}(B[4] + 1, B[4] - 1) = 0$ to the constraint $\rho$. This is equivalent to adding the third and fourth rows in Figure~\ref{fig:overview}(b).
At Lines~10-11,
we add the constraint $B[0]B[1] = 10 \land \fieldnamefun{B[0..1]} = \fieldname{code}$.
This is equivalent to adding the fifth and sixth rows in Figure~\ref{fig:overview}(b).
We regard $B[0]$ and $B[1]$ as a single field as they are used in a single value~$B[0]B[1]$.

Similarly, 
after Line~15,
we merge the paths forked from Line~11 as the constraint $(B[3] = 0 \lor B[3] \ne 0) \land \fieldnamefun{B[2]} = \fieldname{state}$ and append it to the path constraint $\rho$.
This is equivalent to adding the last row in Figure~\ref{fig:overview}(b).
% \noindent
% \begin{minipage}{\columnwidth}
%     \begin{align*}
%         \rho \equiv &~\fieldnamefun{B[4]} = \fieldname{ctrl} \land (B[5] = 0\lor B[5] \ne 0) \\
%         & \land B[6] > 0 \land  \Theta_{\kappa_{3}}(B[4] + 1, B[4] - 1) = 0 \\
%         & \land B[0]B[1] = 10 \land \fieldnamefun{B[0..1]} = \fieldname{code} \\
%         & \land (B[3] = 0 \lor B[3] \ne 0) \land \textup{name}(B[2]) = \fieldname{state} 
%     \end{align*}
% \end{minipage}
% \vspace{2mm}\noindent
After Line~15,
we update the value
$\varname{state} = \Theta_{\kappa_{13}}(B[2] + 1, B[2] - 1)$.
In this example, the variable \varname{state} is simply printed at Line 16 and never used in any if-statements or assertions.
Hence, the merged value of \varname{state} is abstracted away from the final constraint.
% The resulting path constraint $\rho$ is equivalent to the AFG in Figure~\ref{fig:overview}(b). Due to the equivalence, in our inference rules, we direct generate the AFG without computing the path constraint first.
%In the graph, conjunctions are denoted as edges and disjunctions are sibling vertices. Thus, a constraint $a\land (b\lor c)$ is represented as a vertex $a$ with edges to two sibling vertices $b$ and $c$.
%In the graph, we label a vertex by $\kappa_i$ or $\lnot\kappa_i$
%if it contains a constraint produced by taking the true branch or the false at Line $i$.
Observe that the size of AFG is linear size with the number of statements. This is critical to scalability.
\hfill $\Box$

\startreviseblock
\newcommand{\LemmaAbsInt}{Given a program in the language defined in Figure~\ref{fig:design_lang},
    the AFG produced by the abstract interpretation is sound and complete.}
\begin{lemma}
    \LemmaAbsInt
    \label{lemma:abstractinterpretation}
\end{lemma}
\dismissreviseblock

\remark{(1) The proofs of the lemma above and all other lemmas are put in Appendix~\ref{app:proof}.}\remark{(2) The paragraph below is moved to Section~\ref{subsec:impl}}

\delete{\defpar{Loops and Repetitive Fields.}
Loops in a protocol parser
are often used to parse repetitive fields.
We follow existing techniques to analyze loops%~\cite{xie2016proteus,saxena2009loop}.
These techniques are good at inferring repetitive fields and how many times a field repeats.
For example,
for the code below,
we will produce the production $S\rightarrow B[0] B[1]$ with two assertions: \textbf{assert}($B[1] < 5$) and \textbf{assert}(\textsf{repeat}($B[1]) = B[0]$).}

\delete{
    \texttt{j = 0;
        \textbf{while}(j < packet[0]) \textbf{assert}(packet[++j] < 5); }}

\delete{Basically, these loop techniques work in two steps.
First,
they analyze several iterations of a loop.
For instance, they analyze three iterations
and get the results, $j=1 \land B[1]<5$, $j=2 \land B[2] < 5$, and $j=3 \land B[3] < 5$, for each iteration.
Second,
they try to inductively summarize the conditions.
For instance, the above results can be summarized to $j = k \land B[k] < 5$, where $1 \le k \le B[0]$ is an induction variable representing the iteration counter.
Note that not all loops can be summarized as above.
This is an inherent limitation of static analysis.
For irregular loops,
we follow the practice of bounded model checking%~\cite{biere1999symbolic}
to unroll loops. %a fixed number of times. 
While this may introduce unsound results, 
we observe few issues in practice.}

\subsection{Localized Graph Unfolding}
\label{subsec:se}

\noindent
Recall that path sensitivity is needed in localized regions during lifting (Challenge 1 in \S\ref{sec:nutshell}).
Specifically, a code region that requires path sensitivity is identified as follows. {\em If a $\Theta_\kappa$-merged value is later used in some path condition $\kappa'$, the individual combinations of branch outcomes of $\kappa$ and $\kappa'$ need to be analyzed separately.} That is, path sensitivity is needed within the code regions of $\kappa$ and $\kappa'$. On the other hand, many  $\Theta_\kappa$-merged values are not used in any later conditionals, the paths within the code region of $\kappa$ do not need to be enumerated. That is, path sensitivity is not necessary.

\begin{figure}
    \centering
    \includegraphics[width=0.85\columnwidth]{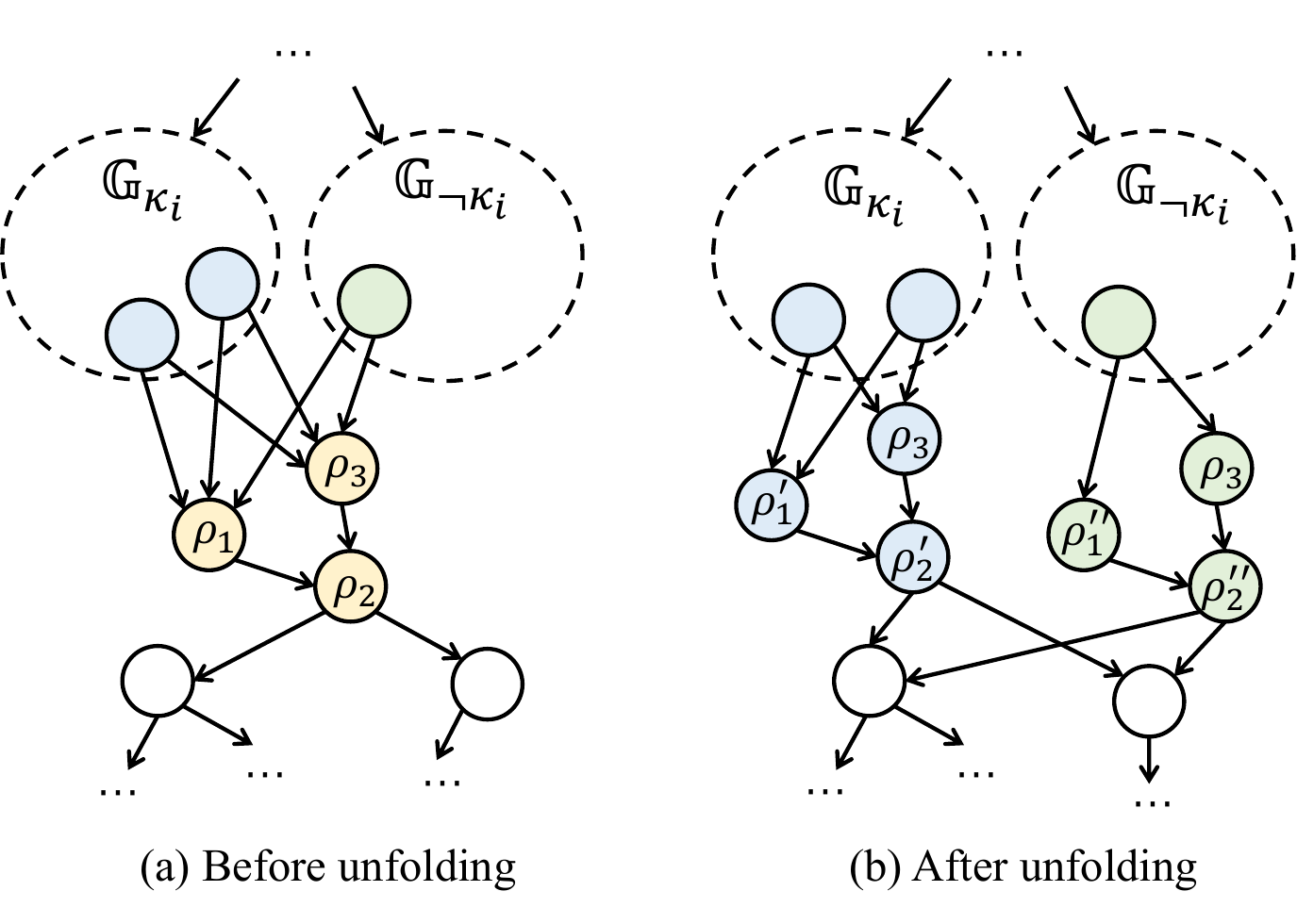}
    \caption{Example of unfolding an AFG.}\label{fig:unfolding}
\end{figure}

Specifically, given an AFG created by the abstract interpretation, we eliminate all $\Theta_\kappa$-merged values by a localized graph unfolding algorithm shown in Algorithm~\ref{alg:unfolding}.
Assume that the AFG to unfold contains a list of $\Theta_\kappa$ operators, e.g., $\Theta_{\kappa_0}$, $\Theta_{\kappa_1}$, and $\Theta_{\kappa_2}$.
The algorithm eliminates $\Theta_{\kappa_i}$ one by one.
For each $\Theta_{\kappa_i}$,
it works in two steps --- slicing (Lines 3-7) and unfolding (Lines 8-11).
To ease the explanation,
we use Figure~\ref{fig:unfolding} for illustration.
In Figure~\ref{fig:unfolding}(a),
without loss of generality,
assume that we are unfolding $\Theta_{\kappa_i}$ in the AFG and that only the constraints $\rho_1$ and $\rho_2$ contain $\Theta_{\kappa_i}$-merged values. 
The exiting vertices of $\mathbb{G}_{\kappa_i}$
and $\mathbb{G}_{\lnot\kappa_i}$ are shown in the figure.

\defpar{Slicing.}
This step delimits the next unfolding step to a local region in AFG.
First, we find all exiting vertices of $\mathbb{G}_{\kappa_i}$
and $\mathbb{G}_{\lnot\kappa_i}$.
We then perform a forward graph traversal (e.g., depth-first search) from the exiting vertices.
Denote the subgraph visited during the traversal as $\mathbb{G}_\textup{forward}$.
Second,
we identify all vertices containing $\Theta_{\kappa_i}$-merged values, e.g., $\rho_1$ and $\rho_2$ in Figure~\ref{fig:unfolding}(a).
A backward graph traversal from them
yields a subgraph denoted as $\mathbb{G}_\textup{backward}$.
The overlapping part of $\mathbb{G}_\textup{forward}$ and $\mathbb{G}_\textup{backward}$, e.g., the yellow part in Figure~\ref{fig:unfolding}(a),
is the graph slice we will perform unfolding, denoted as $\mathbb{G}_\textup{slice}$.
 
\defpar{Unfolding.}
As illustrated in Figure~\ref{fig:unfolding}(b),
we copy the subgraph to unfold, obtaining $\mathbb{G}_\textup{slice}$ and $\mathbb{G}_\textup{slice}'$.
The copy $\mathbb{G}_\textup{slice}$ is connected to $\mathbb{G}_{\kappa_i}$,
and by the definition of the merging operator, all the $\Theta_{\kappa_i}$-merged values are replaced by its first operand.
Similarly, the other copy $\mathbb{G}_\textup{slice}'$ is connected to $\mathbb{G}_{\lnot\kappa_i}$,
and all the $\Theta_{\kappa_i}$-merged values are replaced by its second operand.
Since the subgraphs to unfold are limited in small local regions \revise{in practice}, we significantly mitigate the path-explosion problem, which is sufficient to make our approach scalable.
\revise{Note that we do not claim to have a theoretical bound on the size of subgraphs that need to be unfolded, as path explosion is still an open problem and cannot be completely addressed in theory, similar to all previous path-sensitive  analyzers.}

\newcommand{\LemmaUnfolding}{The unfolded AFG does not contain $\Theta_{\kappa}$-merged values
    and represents an equivalent constraint as the original AFG.}
\begin{lemma}
    \LemmaUnfolding
    \label{lemma:unfolding}
\end{lemma}

\noindent
{\bf Example (continued).}
In Figure~\ref{fig:overview}(b), the value merged by $\Theta_{\kappa_3}$ indicates that the branches forked at Line 3
need a path-sensitive analysis
and delimits the analysis to the local region colored gray.
To distinguish the two branches,
the gray region in (b) is unfolded to two disjoint paths in (c),
which eliminates the $\Theta$-merged values and make the two semantic relations among $B[4]$, $B[5]$, and $B[6]$ explicit: $B[5]=0\land B[6]>0\Leftrightarrow B[4]+1=0$; and $B[5]\ne0\land B[6] > 0\Leftrightarrow B[4]-1=0$.
In contrast, the $\Theta_{\kappa_{13}}$ value in variable {\it state} is never used in any conditional, suggesting that we do not need to unfold the region led by Line 13. \hfill $\Box$

\begin{algorithm}[t]\footnotesize
    \caption{Unfolding.}
    \label{alg:unfolding}
    \SetKwFunction{GenProduct}{unfold}
    \SetKwProg{Proc}{Procedure}{}{}
    \Proc{\GenProduct{$\mathbb{G}$}}{
        \ForEach{\textup{operator} $\Theta_{\kappa_i}$ \textup{in} $\mathbb{G}$}{
            $\mathbb{G}_\textup{forward}\leftarrow$ subgraph reachable from but excluding $\mathbb{G}_{\kappa_i}$ and $\mathbb{G}_{\lnot\kappa_i}$\;
            
            $\mathbb{V}\leftarrow$ all vertices including $\Theta_{\kappa_i}$ expressions\;
            
            $\mathbb{G}_\textup{backward}\leftarrow$ subgraph that can reach any vertex in $\mathbb{V}$, including $\mathbb{V}$\;
            
            $\mathbb{G}_\textup{slice}\leftarrow$ overlapping subgraph of $\mathbb{G}_\textup{forward}$ and $\mathbb{G}_\textup{backward}$\;
            
            $\mathbb{G}_\textup{slice}'\leftarrow$ a copy of $\mathbb{G}_\textup{slice}$, including all its incoming/outgoing edges\;
            
            disconnect $\mathbb{G}_\textup{slice}$ from $\mathbb{G}_{\lnot\kappa_i}$\;
            replace all $\Theta_{\kappa_i}$ expressions in $\mathbb{G}_\textup{slice}$ with their first operands\;
            
            disconnect $\mathbb{G}_\textup{slice}'$ from $\mathbb{G}_{\kappa_i}$\;
            replace all $\Theta_{\kappa_i}$ expressions in $\mathbb{G}_\textup{slice}'$ with their second operands\;
        }
    }
\end{algorithm}

\subsection{Localized Graph Reordering}\label{subsec:spec}

As illustrated in Figure~\ref{fig:overview},
bytes in a packet may not appear in the order in a program path, e.g., $B[5]$ may precede $B[2]$.
To produce legitimate BNF productions, we need to reorder them to produce the ordered AFG.
Then transforming an ordered AFG to BNF productions is straightforward.
%as shown in Figure~\ref{fig:overview},
%this section focuses on the graph reordering algorithm.
We first define the concepts of {\em vertical decomposition} (VD) and {\em horizontal decomposition} (HD).

\begin{figure*}[t]
    \centering
    \includegraphics[width=\textwidth]{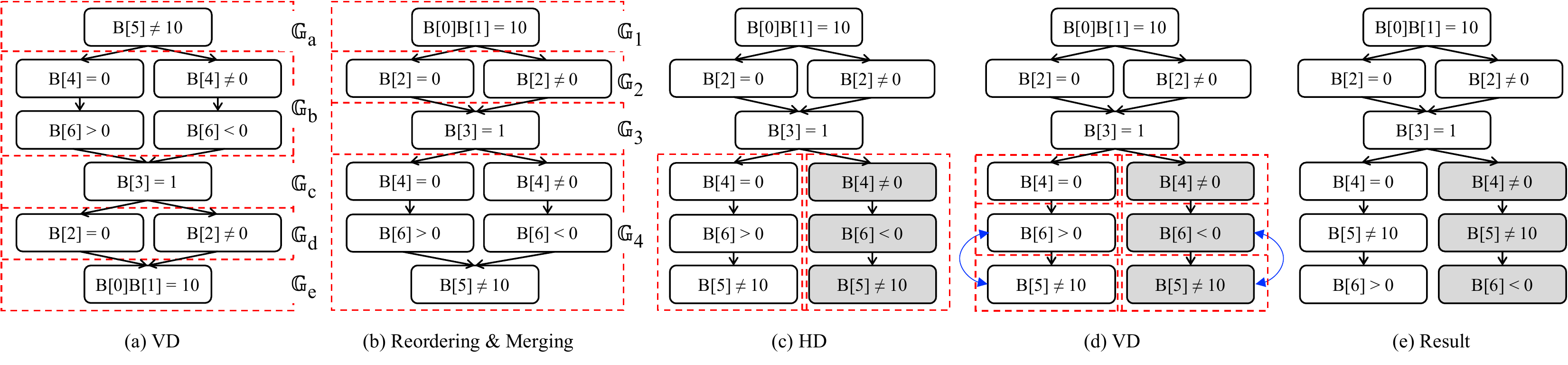}
    \caption{Example of Algorithm~\ref{alg:bnf}.}
    \label{fig:ordering_example}
\end{figure*}

\begin{definition}[VD]
    Given an unfolded AFG $\mathbb{G}=\mathbb{G}_1\bowtie\mathbb{G}_2\bowtie\dots\bowtie\mathbb{G}_n$, namely, the exit vertices in $\mathbb{G}_i$ are fully connected to the entry vertices in $\mathbb{G}_{i+1}$,
    its vertical decomposition is the sequence of subgraphs,
    denoted as $\textsf{VD}(\mathbb{G}) = \mathbb{G}_1\mathbb{G}_2\dots\mathbb{G}_n$.
\end{definition}

\begin{definition}[HD]
    Given an unfolded AFG $\mathbb{G}$, its horizontal decomposition is a set of subgraphs,
    each of which is rooted at a single entry vertex in the AFG and includes the subgraph reachable from the entry vertex,
    denoted as $\textsf{HD}(\mathbb{G}) = \mathbb{G}_1 | \mathbb{G}_2 | \dots | \mathbb{G}_n$.
\end{definition}

\begin{figure}[t]
    \centering
    \includegraphics[width=\columnwidth]{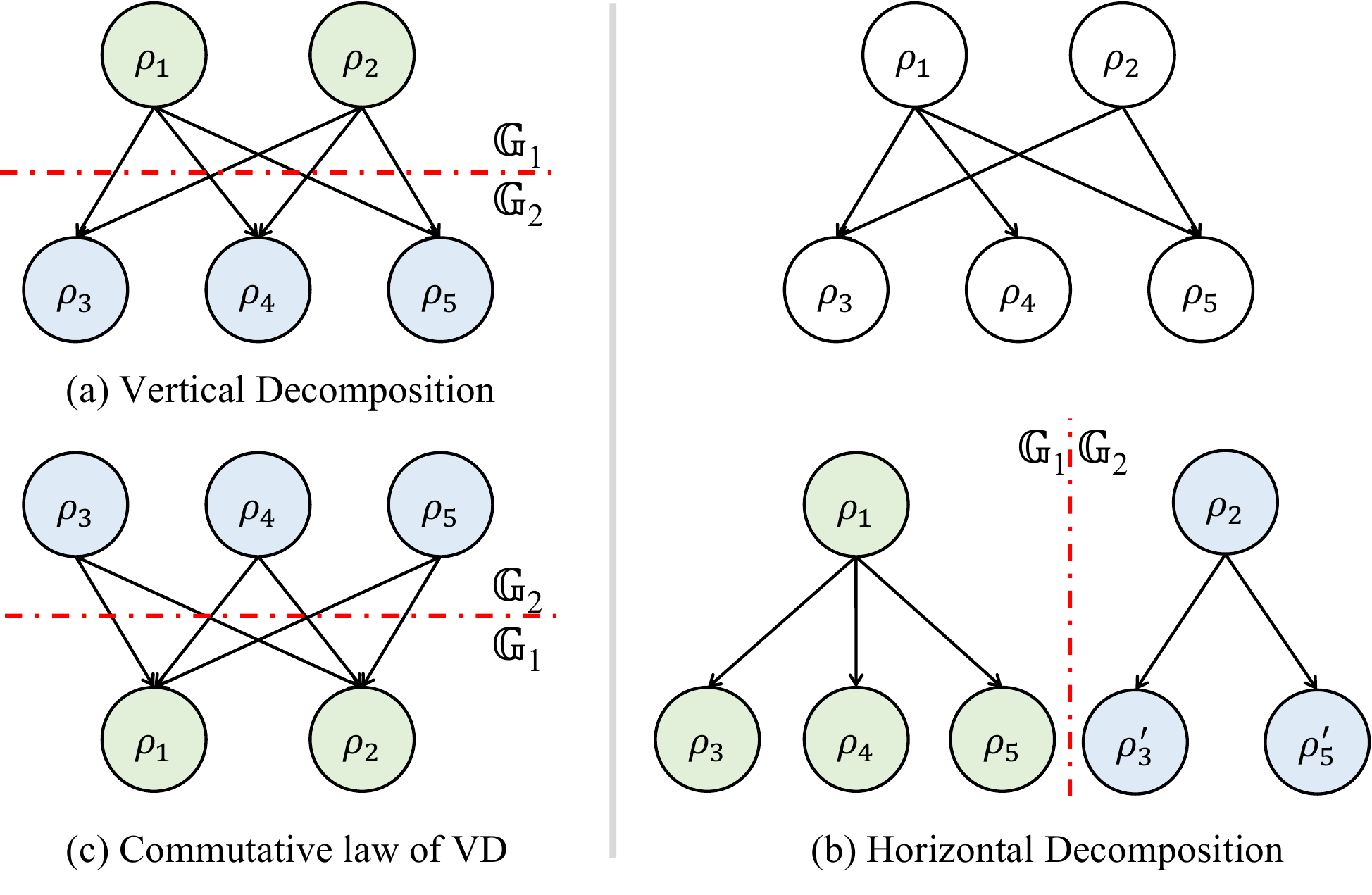}
    \caption{Decomposition for graph reordering.}
    \label{fig:decomposition}
\end{figure}

%\noindent
Figure~\ref{fig:decomposition}(a) shows an example of vertical decomposition,
where the graph is decomposed into two parts, one containing the vertices $\rho_1$ and $\rho_2$,
and the other containing the vertices $\rho_3$, $\rho_4$, and $\rho_5$.
The graph in Figure~\ref{fig:decomposition}(b) cannot be vertically decomposed because 
the upper two vertices are not fully connected to the other three.
Instead, 
it can be horizontally decomposed into two parts,
one containing the vertices $\rho_1$, $\rho_3$, $\rho_4$, and $\rho_5$,
and the other containing the vertices $\rho_2$, $\rho_3'$, and $\rho_5'$.
Here, $\rho_3'$ and $\rho_5'$ are copies of $\rho_3$ and $\rho_5$, respectively.
As illustrated in the example and stated in Lemma~\ref{lemma:decomp},
the AFGs before and after decomposition contain the same number of paths
and the constraint represented by each path is not changed.

\newcommand{\LemmaDecomposition}{AFGs before and after decomposition are equivalent in representing path constraint.}
\begin{lemma}
    \LemmaDecomposition
    \label{lemma:decomp}
\end{lemma}

The decomposition has three properties.
First, the horizontal decomposition is more expensive than the vertical one as it may copy vertices.
Hence, Algorithm~\ref{alg:bnf} always tries the vertical decomposition first.
Second,
as stated in Lemma~\ref{lemma:recursive}, the decomposition can be recursively performed on a graph and its subgraphs.
For instance, after the horizontal decomposition in Figure~\ref{fig:decomposition}(b),
we can further apply vertical decomposition to each subgraph.
This property allows us to describe our reordering approach as a recursive process in Algorithm~\ref{alg:bnf}.
Third, the vertical decomposition follows the commutative law stated in Lemma~\ref{lemma:position}.
For instance, 
after switching $\mathbb{G}_1$ and $\mathbb{G}_2$ in Figure~\ref{fig:decomposition}(a),
we get the graph in Figure~\ref{fig:decomposition}(c), which is equivalent to the original graph
because they represent equivalent path constraints:
$(\rho_1\lor \rho_2) \land (\rho_3\lor \rho_4\lor \rho_5)$
and 
$(\rho_3\lor \rho_4\lor \rho_5) \land (\rho_1\lor \rho_2)$.
Such a commutative property allows us to reorder vertices in Algorithm~\ref{alg:bnf}. 

\newcommand{\LemmaRec}{If an AFG with multiple vertices cannot be vertically decomposed, each subgraph after horizontal decomposition contains a single vertex or can be vertically decomposed.}
\begin{lemma}
    \LemmaRec
    \label{lemma:recursive}
\end{lemma}

\newcommand{\LemmaPos}{Switching the position of subgraphs in VD yields an AFG that represents an equivalent constraint as the original AFG.}
\begin{lemma}
    \LemmaPos
    \label{lemma:position}
\end{lemma}

Algorithm~\ref{alg:bnf} first tries to vertically decompose the input AFG (Line 2). If failed, Lemma~\ref{lemma:recursive} allows us to horizontally decompose it into subgraphs and recursively order each subgraph (Lines 14-15).
If VD succeeds in splitting AFG into a list of subgraphs, these subgraphs are reordered by byte indices (Lines 3-5). 
Figure~\ref{fig:ordering_example}(a) and Figure~\ref{fig:ordering_example}(b) illustrate this step.
In Figure~\ref{fig:ordering_example}(a),
the AFG is vertically decomposed into five subgraphs, $\mathbb{G}_a$, $\mathbb{G}_b$, $\mathbb{G}_c$, $\mathbb{G}_d$, and $\mathbb{G}_e$, which are respectively put in five dashed boxes.
The minimum byte indices of the subgraphs are 5, 4, 3, 2, and 0.
Figure~\ref{fig:ordering_example}(b) shows the AFG after reordering the subgraphs based on the minimum byte indices.
After reordering, since $\mathbb{G}_a$ and $\mathbb{G}_b$ contain overlapping byte indices\footnote{The range of byte indices in $\mathbb{G}_a$ is [5, 5], and the range in $\mathbb{G}_b$ is [4, 6]. The former is a subset of the latter. Thus, they overlap each other.}, 
they are merged into a single subgraph, i.e., $\mathbb{G}_4$ in Figure~\ref{fig:ordering_example}(b).
In this example, the subgraphs after reordering and merging are put in the array $\mathcal{A} = [ \mathbb{G}_1, \mathbb{G}_2, \mathbb{G}_3, \mathbb{G}_4 ]$.
These subgraphs are ordered and contain mutually exclusive byte indices.

\begin{algorithm}[t]\footnotesize
    \caption{Reordering.}
    \label{alg:bnf}
    \SetKwFunction{GenProduct}{reorder}
    \SetKwFunction{VD}{VD}
    \SetKwFunction{HD}{HD}
    \SetKwProg{Proc}{Procedure}{}{}
    \Proc{\GenProduct{$\mathbb{G}$}}{
        \If {$\VD(\mathbb{G})=\mathbb{G}_a\mathbb{G}_b\dots$} {
            reorder $\mathbb{G}_a$, $\mathbb{G}_b$, $\dots$, based on the min byte index of each subgraph\;
            merge adjacent subgraphs if they contain overlapping byte indices\;
            \textbf{let} $\mathcal{A}\leftarrow$ [$\mathbb{G}_1$, $\mathbb{G}_2$, $\dots$] be subgraphs after reordering and merging\;
            
            \ForEach{$\mathbb{G}_i\in \mathcal{A}$}{
                \If {$\mathbb{G}_i=\mathbb{G}_{i,1}\bowtie\mathbb{G}_{i,2}\bowtie\dots$ \textup{is a merged graph}}{
                    \textbf{assume} $\mathbb{G}_{i,j}$ has multiple entry vertices\;
                    switch the position of $\mathbb{G}_{i,j}$ and $\mathbb{G}_{i,1}$ in $\mathbb{G}_i$\;
                    \textbf{assume} $\HD(\mathbb{G}_i) = \mathbb{G}_{a'} | \mathbb{G}_{b'} | \dots $\;
                    $\GenProduct(\mathbb{G}_{a'});~\GenProduct(\mathbb{G}_{b'});~\dots$\;
                }
                \Else {
                    $\GenProduct(\mathbb{G}_{i})$\;
                }
            }
        }
        \ElseIf{$\HD(\mathbb{G}) = \mathbb{G}_a | \mathbb{G}_b | \dots $}{
            $\GenProduct(\mathbb{G}_a);~\GenProduct(\mathbb{G}_b);~\dots$\;
        }
        \textbf{else }~\tcp{a single-vertex graph, do nothing}
    }
    \vspace{1.5mm}
\end{algorithm}

\begin{algorithm}[t]\footnotesize
    \caption{Packet Format in BNF.}
    \label{alg:bnf2}
    \SetKwFunction{GenProduct}{bnf}
    \SetKwFunction{VD}{VD}
    \SetKwFunction{HD}{HD}
    \SetKwProg{Proc}{Procedure}{}{}
    \Proc{\GenProduct{$\mathbb{G}$}}{
        $L\leftarrow$ new non-terminal symbol\;
        \If {$\VD(\mathbb{G})=\mathbb{G}_a\mathbb{G}_b\dots$} {
            $L\rightarrow\GenProduct(\mathbb{G}_a)~\GenProduct(\mathbb{G}_b)~\dots$\;
        }
        \ElseIf{$\HD(\mathbb{G}) = \mathbb{G}_a | \mathbb{G}_b | \dots $}{
            $L\rightarrow\GenProduct(\mathbb{G}_a)~|~\GenProduct(\mathbb{G}_b)~|~\dots$\;
        }
        \Else{
            \tcp{a single vertex containing $B[i],B[i+1],\dots, B[i+k]$} 
            $L\rightarrow B[i]B[i+1]\dots B[i+k]$ with assertions in the vertex\;
        }
        \Return $L$\;
    }
    \vspace{1.5mm}
\end{algorithm}

We then recursively reorder subgraphs in $\mathcal{A}$ (Lines~6-13).
Especially, for a merged subgraph, e.g., $\mathbb{G}_4=\mathbb{G}_a\bowtie\mathbb{G}_b$ in the example,
since we have tried vertical decomposition, which does not work as neither $\mathbb{G}_a\ \mathbb{G}_b$ nor $\mathbb{G}_b\ \mathbb{G}_a$ respects the stream order,
we turn to horizontal decomposition (Lines~8-11).
Lines~8-9 ensure the feasibility of horizontal decomposition and
Line~10 performs the decomposition.
Figure~\ref{fig:ordering_example}(c) illustrates this step,
where the subgraph $\mathbb{G}_4$
is horizontally decomposed into the white and the gray parts.
Each part then is recursively reordered (Line~11).
Figure~\ref{fig:ordering_example}(d) shows that
the white and the gray parts are recursively split by vertical 
decomposition and reordered as indicated by the arrows,
yielding the ordered AFG in Figure~\ref{fig:ordering_example}(e).
Lemma~\ref{lemma:ordering} states the correctness of Algorithm~\ref{alg:bnf}.

\newcommand{\LemmaOrdering}{Algorithm~\ref{alg:bnf} yields an ordered AFG, which represents an equivalent constraint as the input AFG.}
\begin{lemma}
    \LemmaOrdering
    \label{lemma:ordering}
\end{lemma}

\noindent\textbf{From Ordered AFG to BNF-like Format.}
It is straightforward to translate an ordered AFG to packet formats in BNF.
Due to its simplicity, the detailed discussion is elided and the formal algorithm is put in Algorithm~\ref{alg:bnf2}.
As an example, Figure~\ref{fig:overview}(e) shows the inferred packet format
where
$S$ is the start symbol that represents the whole graph and each non-terminal $L_i$ represents a subgraph ---
$L_1$ represents the path prefix containing $B[0]$, $B[1]$, and $B[2]$;
$L_2$ and $L_3$ represent two possible constraints of $B[3]$;
and $L_4$ and $L_5$ stand for the two path suffixes containing $B[4]$, $B[5]$, and $B[6]$.

\startreviseblock
\subsection{Soundness and Completeness in Practice}
\label{subsec:impl}

As proved in Appendix~\ref{app:proof},
Lemmas~\ref{lemma:afg}-\ref{lemma:ordering} together guarantee the theoretical soundness and completeness of our approach
for a program written in our abstract language.
In practice, we need to handle common program structures not included in the abstract language,
such as function calls, pointers, and loops.
This section discusses how we handle them in our implementation and their
effects on soundness or completeness.

\defpar{Pointers.}
In the previous discussion, we focus on building an AFG for format inference. Pointer operations are not directly related to AFG. 
In the implementation, we follow existing works~\cite{sui2011spas} to resolve pointer relations,
which helps us identify what values may be loaded from a memory location.
For instance,
when visiting an assertion in the program such as \textsf{{assert}(*(p + 1) > 1)} where \texttt{p} is a pointer,
if the pointer analysis tells us \texttt{p+1} points to a memory location storing the value $B[5]$ on the condition $\rho$,
we then compute and include the constraint $\rho \Rightarrow B[5] > 1$ (which equals $\lnot \rho \lor B[5] > 1$) in AFG.
Pointer operations such as \texttt{p+1} are not a part of path constraints and, thus, are not included in AFG.
That is, according to the assertion rule in Figure~\ref{fig:semantics_basic} and assuming the 
AFG before the assertion is $\mathbb{G}$,
the AFG after the assertion is $\mathbb{G}\bowtie \textsf{AFG}(\lnot \rho \lor B[5] > 1)$.
Since the pointer analysis we use is sound and path-sensitive,
it allows \tool\ to be sound and highly precise.

\defpar{Function Calls.} Although we do not include function calls in our abstract language for simplicity, 
our system is inter-procedural as a call statement is equivalent to a list of assignments from the actual parameters
to the formals, and a return statement is an assignment from the return value to its receiver.
Thus, in our analysis, function calls and returns are treated as assignments. This treatment does not degrade soundness and completeness.
Especially, for recursive function calls, we convert them to loops, which are discussed below.
\dismissreviseblock

\defpar{Loops and Repetitive Fields.}
Loops in a protocol parser are often used to parse repetitive fields~\cite{cui2008tupni}.
We follow existing techniques to analyze loops~\cite{xie2016proteus,saxena2009loop},
which are good at inferring repetitive fields and how many times a field repeats.
For example,
\delete{for }the code below \revise{parses a packet where $B[0]$ represents the packet length and contains a positional constraint that all bytes after $B[0]$ are less than five. For this example},
we produce the production $S\rightarrow B[0] B[1]$ with two semantic constraints: $B[1] < 5$ and $\textsf{repeat}(B[1]) = B[0]$.

\ffbox{
    \begin{minipage}{0.95\columnwidth}\centering
        \small\noindent
        \textsf{j = 0;\enskip\enskip
            \textbf{while}(j < packet[0])\enskip\{\enskip \textbf{assert}(packet[++j] < 5); \enskip\} }
    \end{minipage}
}

\noindent Basically, the loop analysis works in two steps.
First,
they analyze several iterations of a loop.
For instance, it analyzes three iterations
and gets the results, $j=1 \land B[1]<5$, $j=2 \land B[2] < 5$, and $j=3 \land B[3] < 5$, for each iteration.
Second,
it inductively infers the conditions.
For instance, the above results can be inductively summarized to $j = k \land B[k] < 5$, where $1 \le k \le B[0]$ is an induction variable representing the iteration counter.
\revise{Equivalently, we write the constraint as $B[1] < 5 \land \textsf{repeat}(B[1]) = B[0]$
    and, by the definition of AFG, represent it as an edge from a vertex containing $B[1] < 5$ to one containing $\textsf{repeat}(B[1]) = B[0]$. If the inductive inference succeeds, its result is sound and complete.}

\startreviseblock\vspace{1mm}
\noindent
\textbf{Irregular Loops.} 
Nevertheless, not all loops can be inductively summarized as above.
This is an inherent limitation of static analysis.
For irregular loops,
we follow the practice of bounded model checking~\cite{biere1999symbolic} to unroll loops a fixed number of times.
While this may introduce unsoundness and incompleteness,
we observe few irregular loops (involved in packet parsing) in practice and their influence is limited.
In other words,
it may produce some unsound or incomplete constraints for some fields in a packet
but such unsoundness and incompleteness are rarely propagated to other fields. 

\begin{figure}[t]
    \centering
    \includegraphics[width=\columnwidth]{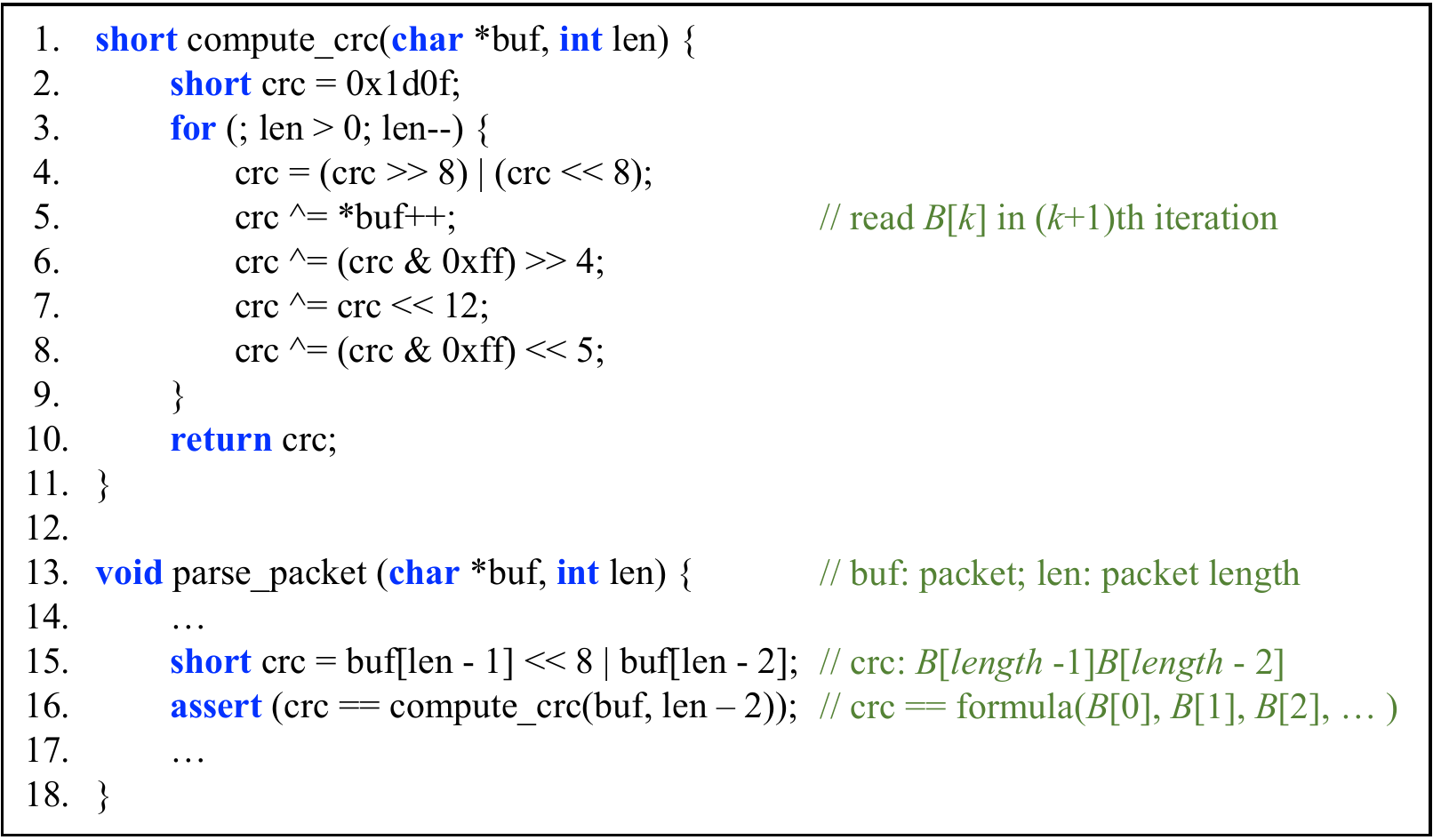}
    \caption{An irregular loop that computes checksum.}
    \ifdiff \else \vspace{-1.5mm}\fi
    \label{fig:cksum}
\end{figure}

The most common irregular loops related to protocol formats are those computing the checksum.
Figure~\ref{fig:cksum} shows an example
where the function \textit{compute\_crc} computes the CRC value as the checksum
and the function \textit{parse\_packet} compares the computed CRC value to the received one in the packet.
The CRC value is computed by a loop, which cannot be inductively summarized.
This is because, unlike the variable $j$ in the previous example, 
we cannot write the value of \textit{crc} as a formula parameterized by the loop counter~$k$.

Thus, we unroll the loop a fixed number of times $t$, yielding an unsound and incomplete formula of \textit{crc} that relies on $B[0]$, $B[1]$, $\dots$, $B[t-1]$, denoted as $f(B[0], B[1], \dots, B[t-1])$.
At Line~16, this formula is compared to the CRC field in the packet.
Generally, by the assertion rule in Figure~\ref{fig:semantics_basic},
an AFG node with the constraint $B[\textit{length}-1]B[\textit{length}-2]=f(B[0], B[1], \dots, B[t-1])$
should be appended to the AFG.
However, in the implementation, we skip such unsound and incomplete constraints,
so that they are not included in the protocol format output by \tool.
Excluding these constraints keeps our inferred format sound but may lose some precision (i.e., completeness) as we miss some constraints.

While we cannot compute precise non-recursive FOL constraints for checksum fields (this is a limitation shared by existing works on protocol format reverse engineering),
as discussed in \S\ref{subsec:ai}, we can still infer the boundary of checksum fields. 
Such field boundaries, together with all sound constraints we compute,
can already support many security applications (see \S\ref{subsec:eval_app}).
\dismissreviseblock

%% file: evaluation.tex
\section{Evaluation}
\label{sec:eval}

\noindent
We implement our method as a tool, namely \tool, to lift packet formats from source code in C.
It is implemented on top of the LLVM (12.0.0) compiler infrastructure~\cite{lattner2004llvm} and the Z3 (4.8.12) SMT solver~\cite{de2008z3}.
The source code of a protocol is compiled into the LLVM bitcode, where we perform our static analysis.
In the analysis, Z3 is used to represent abstract values as symbolic expressions
and compute/solve path constraints.
All experiments are run on a Macbook Pro (16-inch, 2019)
equipped with an 8-core 16-thread Intel Core i9 CPU with 2.30GHz speed and 32GB of memory.

As shown in Table~\ref{tab:bench}, we have run \tool\ over a number of protocols.
They are from different codebases (e.g., Linux and LWIP) and domains (e.g., IoT and routing).
They include widely-used ones such as TCP/IP
and niche ones like APDU that is used in smart cards.
As shown in the table, 
\revise{the size of the code involved in a protocol parser ranges from 3KLoC to 59KLoC, and}
it takes \tool\ <1min to infer \mreplace{protocol formats}{the format of each protocol}. 
Determining the precision and recall of the inferred formats requires manually comparing them with their official documents.
We cannot afford to manually inspect all protocols because
we have to learn a lot of domain-specific knowledge to understand a protocol, which is time-consuming and not very related to our core contribution to the static analysis.
In the remaining experiments,  
we focus on the first ten, which are from different codebases.
We believe that other protocols in the same codebases are implemented in similar manners and, thus, do not introduce extra challenges.
We use these protocols/codebases because of two reasons.
First, their repositories in GitHub are relatively active,
which makes it easy to get feedback from developers~when we report bugs.
Second, they have their own fuzzing drivers, meaning~that they have been extensively fuzzed by the developers themselves.
Thus, their code is expected to be of high quality and an approach that can find vulnerabilities in their codebase is highly effective.

\begin{table}[t]\scriptsize
    \centering
    \caption{Protocols and Their Codebases for Evaluation}
    \label{tab:bench}
    \begin{tabular}{cp{16mm}ccp{33mm}}
        \toprule
        \multirow{2}{*}{\textbf{Name}} & \multirow{2}{*}{\textbf{Codebase}} & \textbf{Size} & \textbf{Time} & \multirow{2}{*}{\textbf{Description}}\\
        &     & (kloc) & (sec.)                    &                                         \\
        \midrule
        L2CAP    & linux/bluetooth~\cite{linux}& 38   &12& logical link ctrl and adaptation proto.         \\
        SMP      &  linux/bluetooth~\cite{linux} & 12   &2& low energy security manager proto.                 \\%\midrule
        APDU     & opensc~\cite{opensc}        &3   &3             & application proto. data unit                       \\%\midrule
        OSDP     & libosdp~\cite{libosdp}        &14  &27           & open supervised device proto.                      \\%\midrule
        SSQ      & libssq~\cite{libssq}              & 8   & 1   & source server query proto.                         \\%\midrule
        TCP/IP   & lwip~\cite{lwip}                    &41  &53& transport control \& internet proto.         \\
        IGMP/IP  & lwip~\cite{lwip}                   &17  &16& internet group mgmt. \& internet proto. \\%\midrule
        QUIC     & ngtcp2~\cite{ngtcp2}         &59  &  11  & general-purpose transport layer proto.   \\%\midrule
        BABEL    & frrouting~\cite{frrouting}   &7  & 9& a distance-vector routing proto.                   \\
        IS-IS    &   frrouting~\cite{frrouting}    &22 &6& intermediate system (IS) to IS proto. \\
        \midrule
        A2MP&linux/bluetooth~\cite{linux}&16&2&amp manager proto.\\
        BNEP&linux/bluetooth~\cite{linux}&15&3&BT network encapsulation proto.\\
        CMTP&linux/bluetooth~\cite{linux}&20&1&c-api message transport proto.\\
        HIDP&linux/bluetooth~\cite{linux}&17&4&human interface device proto.\\
        UDP&lwip~\cite{lwip}&37&33&user datagram proto.\\
        ICMP&lwip~\cite{lwip}&22&12&internet control message proto.\\
        DHCP&lwip~\cite{lwip}&25&43&dynamic host configuration proto.\\
        ICMP6&lwip~\cite{lwip}&30&54&internet control message proto. v6\\
        DHCP6&lwip~\cite{lwip}&35&51&dynamic host configuration proto. v6\\
        BGP&frrouting~\cite{frrouting}&13&2&border gateway proto.\\
        LDP&frrouting~\cite{frrouting}&20&5&label distribution proto.\\
        BFD&frrouting~\cite{frrouting}&10&17&bidirectional forwarding detection\\
        VRRP&frrouting~\cite{frrouting}&8&12&virtual router redundancy proto.\\
        EIGRP&frrouting~\cite{frrouting}&14&21&interior gateway routing proto. \\
        NHRP&frrouting~\cite{frrouting}&11&11&next hop resolution proto.\\
        OSPF2&frrouting~\cite{frrouting}&9&14&open shortest path first v2\\
        OSPF3&frrouting~\cite{frrouting}&7&16&open shortest path first v3\\
        RIP1&frrouting~\cite{frrouting}&11&13&routing information proto. v1\\
        RIP2&frrouting~\cite{frrouting}&11&15&routing information proto. v2\\
        RIPng&frrouting~\cite{frrouting}&7&41&routing information proto. for ip6\\
        \bottomrule
    \end{tabular}
    \ifdiff \else \vspace{2.5mm}\fi
\end{table}

% \defpar{Research Questions.}
% We focus on three research questions.
% First, regarding the technical contribution,
% we study if our three-step design effectively mitigates path explosion for format inference (\S\ref{subsec:eval_klee}).
% Second,
% we evaluate the precision and recall of the inferred packet formats, compared to four state-of-the-art baselines (\S\ref{subsec:eval_precision}).
% Third,
% we evaluate how our inferred packet formats can help three security applications, including protocol fuzzing,
% network traffic auditing, and network intrusion detection (\S\ref{subsec:eval_app}).

\subsection{Effectiveness of the Three-Step Design}
\label{subsec:eval_klee}

\begin{figure}[t]
    \centering
    \includegraphics[width=\columnwidth]{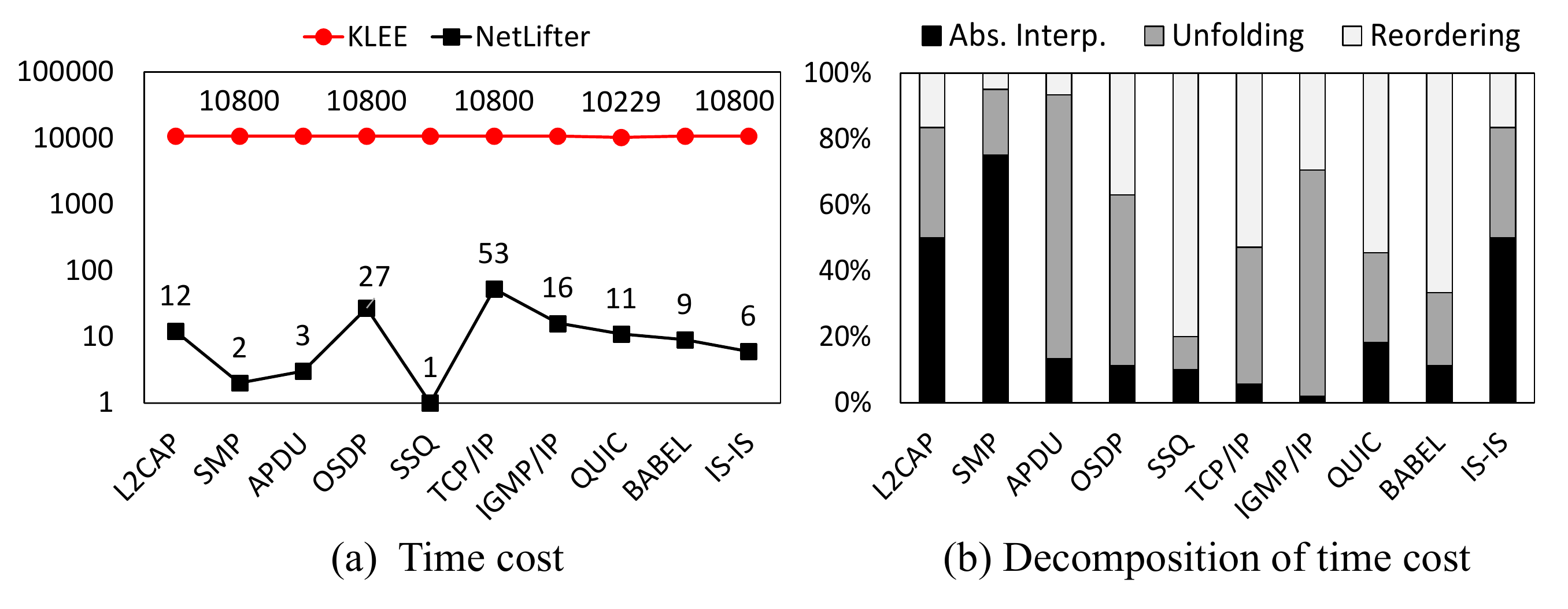}
    \caption{Time cost (seconds) and its decomposition.}
    \label{fig:klee}
\end{figure}

For technical contributions,
we explained in \S\ref{sec:nutshell} that our static analysis avoids individually exploring program paths
to address two challenges.
To show the importance of our solution,
we implement a baseline that employs a well-known symbolic executor, KLEE~\cite{cadar2008klee}, to infer packet formats.
Similar to our solution, it infers packets by computing path constraints.
Different from our solution, it has to analyze individual program paths.
We then compare their time cost of format inference.
The results are shown in Figure~\ref{fig:klee}(a) in log scale.
The line chart shows that the KLEE-based approach runs out of time ($\ge3$ hours) for almost all protocols.
\revise{We use a three-hour time budget here as it is sufficient to show the advantage of our approach over symbolic execution.} \mreplace{In contrast, we}{As plotted in Figure~\ref{fig:klee}(a), \tool} can finish in one minute.
Figure~\ref{fig:klee}(b) shows the decomposition of \tool's time cost,
indicating that the three steps of \tool\ respectively take 14\%, 44\%, and 42\% of the total time.

\subsection{Precision and Recall of Packet Formats}
\label{subsec:eval_precision}

As discussed in \S\ref{sec:intro},
existing techniques focus on network trace analysis (category one) or dynamic program analysis (category two).
We refer to both of them as dynamic analyses as they rely on dynamically captured network packets as their inputs.
We cannot find any static program analysis that infers formats from a protocol parser.
Thus, while the dynamic analyses have a different assumption from our static analysis,
not for a comparative purpose but to show the value of our approach,
we evaluate \tool\ with two network trace analyses, i.e.,
NemeSys~\cite{kleber2018nemesys,kleber2020message} and NetPlier~\cite{ye2021netplier},
and two dynamic program analyses,
i.e., AutoFormat~\cite{lin2008automatic} and Tupni~\cite{cui2008tupni}.
NemeSys and NetPlier are open-source software and we directly use their implementation.
AutoFormat and Tupni are not publicly available.
We implement them on top of LLVM based on their papers.
We cannot find other open-source dynamic program analyses for evaluation.
We evaluate them in terms of precision and recall.
Given a set of packets, the precision is
the ratio of correctly inferred fields in the packets to all inferred fields.
The recall is the ratio of correctly inferred fields to all fields in the ground truth.
\revise{To compute the precision and recall,
we manually build the formats based on the protocols' official documents or source code.
We then write scripts to compare the inferred and the manually-built formats.}

\defpar{Dynamic Analysis.}
To use dynamic analyses, 
we follow their original works to collect 1000 network packets for each protocol 
from publicly available datasets~\cite{packetcapture,packetcapture2,ring2019survey}.
% When there are not 1000 packets available in the public datasets,
% we randomly generate valid packets based on the formats produced by our static analysis.
% To some degree, this situation of lacking input packets
% confirms the value of our static analysis: it is a promising alternative to dynamic analysis especially when inputs are not available.
Table~\ref{tab:precision_recall} shows the precision and recall of the inferred field boundaries.
Network trace analyses often exhibit low precision (<$50$\%) and recall (<$50$\%), because they 
use statistical approaches to align message fields while statistical approaches are known to have inherent uncertainty and their effectiveness heavily hinges on the quality of input packets.

The two dynamic program analyses, especially Tupni, significantly improve the precision due to the analysis of control/data flows in the code.
AutoFormat has a relatively low precision because
it tracks coarse-grained control/data flows.
For instance, AutoFormat regards consecutive bytes of a packet processed in the same calling context as a single field.
However, it is common for a parser to process multiple fields in the same calling context.
Tupni tracks more fine-grained control/data flows, such as predicates in the code,
and, thus, exhibits a higher precision.
As acknowledged by Tupni itself,
it may also produce false fields in many cases.
For instance,
when the value of a multi-byte field is computed by multiple instructions over every single byte in the field,
it will incorrectly split the field into multiple fields.
Despite the high precision achieved by Tupni,
the key problem of these dynamic analyses is their coverage (i.e., recall),
which is often lower than 50\% and may compromise downstream security analyses as discussed in the next subsection.

\revise{Note that simply combining the results of multiple tools does not help improve the quality of the inferred formats.
This is because, when combining the formats inferred by multiple tools,
with the increase of correctly inferred fields, the number of incorrect fields also increases.
For instance, after combining the results of the four dynamic tools,
the precision for OSDP is 0.43, which is even worse than the result when using Tupni independently.
The combined results are shown in the last column of Table~\ref{tab:precision_recall}.}

\begin{table}[t]\scriptsize
    \centering
    \caption{Precision(\%)/Recall(\%) of \underline{Field Boundaries}.}\label{tab:precision_recall}
    \ifdiff \else \vspace{1mm} \fi
    \begin{tabular}{c|c||cc|cc|c}
        \toprule
        {\textbf{Protocol}}& {\tool} & {NemeSys}   & {NetPlier}  & {AutoFormat}  & {Tupni} & Combined \\
        \midrule
        L2CAP&96/98    &14/9    &14/15  &72/32      & 88/41       & 66/49\\ %
        SMP&100/100     &27/37   &20/52  &100/52   & 96/78     & 45/82\\ %
        APDU&100/100   &52/21 &43/45  &44/25        & 100/61  & 58/71\\ %
        OSDP&100/100   &17/11  &10/16  &74/31      &  89/47       & 43/52 \\ %
        SSQ&100/100    &25/1    &26/11  &88/19      &  95/54      & 41/67\\ %
        TCP/IP&98/95    &5/7       &4/12   &24/19      &  88/21     & 39/25\\
        IGMP/IP&99/98  &13/12   &13/22  &35/22     &  92/25     & 54/36\\
        QUIC&97/99      &19/14   &18/25  &70/29     & 86/43       & 69/53\\ %
        BABEL&99/99    &28/14  &37/8   &43/16          & 80/24   & 40/28\\ %
        IS-IS&98/99      &23/5       &19/14  &100/34    & 87/21     & 62/42\\ %
        \bottomrule
    \end{tabular}
    \ifdiff \else \vspace{1mm} \fi
\end{table}

\begin{table}[t]\scriptsize
    \centering
    \caption{Precision(\%)/Recall(\%) of \underline{Field Names}.}\label{tab:names}
    \ifdiff \else \vspace{1mm} \fi
    \begin{tabular}{ccccc}
        \toprule
        L2CAP  & SMP    & APDU    & OSDP    & SSQ      \\
        100/97 & 100/96 & 100/100 & 100/100 & 100/100 \\\midrule
        TCP/IP & IGMP/IP & QUIC  & BABEL  & IS-IS  \\
        100/95 & 100/98  & 96/96 & 100/95 & 100/94\\
        \bottomrule
    \end{tabular}
    \ifdiff \else \vspace{1mm} \fi
\end{table}

\defpar{Static Analysis.}
Table~\ref{tab:precision_recall} shows that, in terms of field boundaries,
our inferred formats cover >$96\%$ formats and produce <$4\%$
false ones.
For many of them, we can produce absolutely correct formats.
\revise{We also miss some fields and report some false ones due to the inherent limitations of static analysis (see \S\ref{subsec:limitations}). These limitations, e.g., the incapability of handling inline-assembly in the source code, will let us lose information during the static analysis, thereby leading to false formats.}
Table~\ref{tab:names} also shows the quality of the inferred~field names.
A name is considered to be correct if it is the same as~the official documents or a reasonable abbreviation, e.g., \fieldname{len} vs. \fieldname{length}.
Overall, we can infer >$94\%$ field names with a precision >$96\%$.
The names provide high-level semantics and help us identify special fields to facilitate security applications as discussed next.

\ifdiff \else \newpage \fi

\subsection{Security Applications}
\label{subsec:eval_app}

\begin{figure}[t]
    \centering
    \includegraphics[width=0.9\columnwidth]{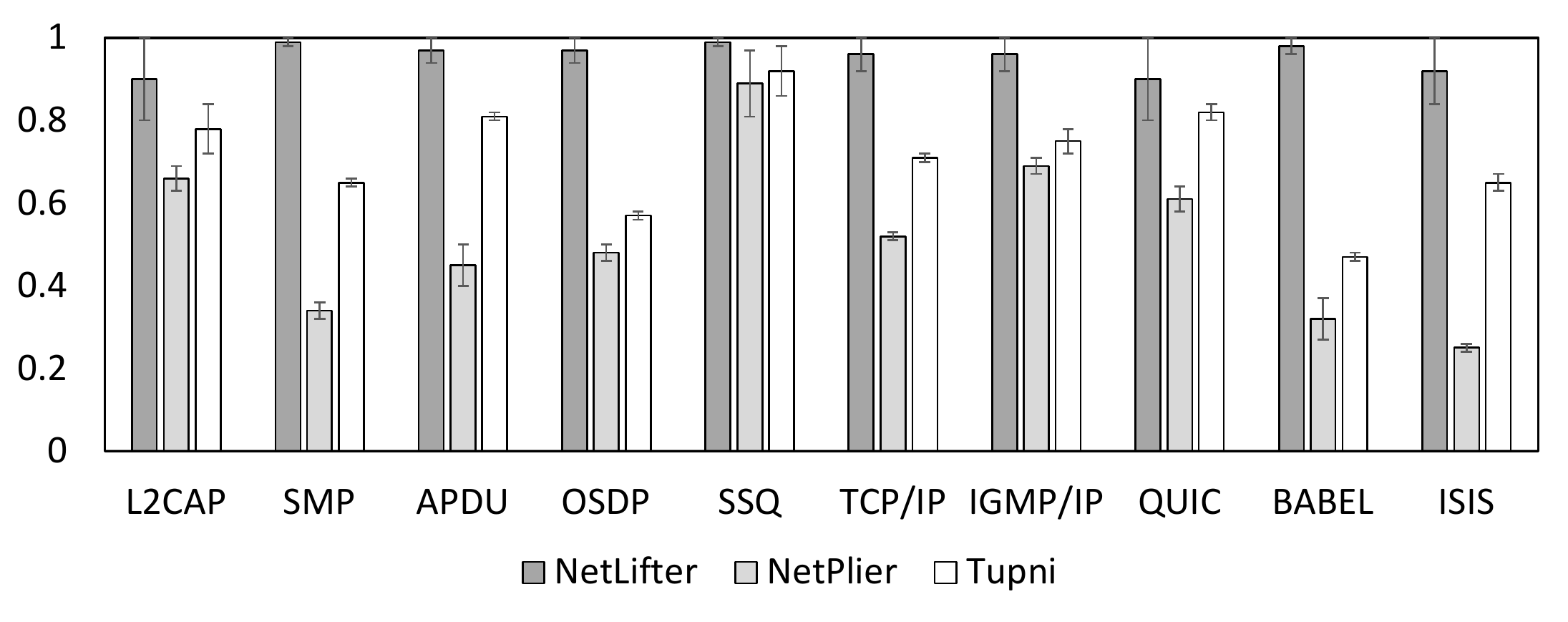}
    \caption{The y-axis is the number of covered branches normalized to one. It shows the branch coverage averaged over twenty runs with a 95\% confidence interval.}\label{fig:fuzz}
\end{figure}

\noindent
\textbf{Protocol Fuzzing.}
To show the value of our approach,
we respectively input the formats inferred by \tool, NetPlier, and Tupni to a typical grammar-based (i.e., format-based) protocol fuzzer, namely BooFuzz~\cite{boofuzz,sulley}.
Particularly, since we can locate checksum fields by names such as \fieldname{checksum} and \fieldname{crc},
in the fuzzing experiments,
we can skip the checksum checks in the code.
This is critical for fuzzing as random mutations in fuzzing can easily invalidate the checksum values~\cite{wang2010taintscope}.
The experiments are performed on a three-hour budget and repeated twenty times to avoid random factors.
\revise{We use a three-hour budget because we observe that the baseline fuzzers rarely achieve new coverage after three hours.} 

The results are shown in Figure~\ref{fig:fuzz}.
Since \tool\ can provide formats with precise field boundaries and semantic constraints,
\tool-enhanced BooFuzz achieves $1.2\times$ to $3.6\times$ coverage compared to others.
\tool-enhanced BooFuzz also detected 53 zero-day vulnerabilities while the others detect only 12. 
All detected vulnerabilities are exploitable as they can be triggered via crafted network packets.
To date, 47 of them have been assigned CVEs.
We can detect more bugs as our inferred formats are of both high precision and high coverage.
In Appendix~\ref{app:bug}, we provide \revise{more details about the fuzzing experiments and the detected bugs}\delete{ a case study of one vulnerability we discovered}.

\defpar{Traffic Auditing and Intrusion Detection.}
Appendix~\ref{app:wireshark} provides an extended study,
where we use the formats inferred by \tool\ and the best baseline, Tupni, to enhance Wireshark and Snort.
We conclude that the precise and high-coverage formats inferred by us are
critical for auditing traffic and detecting intrusions.

%% file: relwork.tex
\section{Related Work}
\label{sec:relwork}

\noindent
Existing techniques that infer packet formats are mainly based on dynamic analysis.
We discuss some typical ones in what follows.
For a broader overview, we refer readers to four surveys~\cite{narayan2015survey,sija2018survey,duchene2018state,li2011survey}.

\defpar{Network Trace Analysis (NTA).} NTA uses statistical methods to identify~field boundaries based on runtime network packets.
%PIP~\cite{pip} uses a bioinformatics approach to align the byte stream of network packets.
Discoverer~\cite{cui2007discoverer}
relies on a recursive clustering approach to recognize packets of the same type.
Biprominer~\cite{wang2011biprominer} uses the variable length pattern to locate protocol keywords
and is enhanced by
ProDecoder~\cite{wang2012semantics}.
AutoReEngine~\cite{luo2013position} uses data mining to identify protocol keywords, based on which packets are classified into different types. 
ReverX~\cite{antunes2011reverse} uses a speech recognition algorithm to identify delimiters in packets.
NemeSys~\cite{kleber2018nemesys,kleber2020message} interprets 
binary packets as feature vectors
and applies an alignment and clustering algorithm to determine the packet format.
NetPlier~\cite{ye2021netplier}
leverages a probabilistic analysis to determine the keyword field,
clusters packets based on the keyword values, and applies multi-sequence alignment to derive packet format.
These techniques do not analyze code and, thus, are different from ours.

\defpar{Dynamic Program Analysis (DPA).} DPA
can be used over both source and binary code.
They work by running protocol parsers against network packets and monitoring runtime control/data flows.
Polyglot~\cite{caballero2007polyglot} uses dynamic taint analysis to infer fixed or variable length fields.
AutoFormat~\cite{lin2008automatic} approximates the field hierarchical structure by monitoring call stacks.
This approach then is extended to both bottom-up and top-down hierarchical structures~\cite{lin2010reverse}.
Wondracek et al.~\cite{wondracek2008automatic} identify delimiters and length fields within a hierarchical structure.
Tupni~\cite{cui2008tupni} tracks fine-grained taint flows to identify packet fields. It also applies loop analysis to infer repeated fields and records path constraints to infer length or checksum fields. 
%Tupni looks similar to our approach but is a dynamic method that works under different assumptions and, hence, addresses different challenges.
ReFormat~\cite{wang2009reformat} recognizes
encrypted fields based on the observation that encrypted fields are processed by a high percentage of arithmetic/bitwise instructions.
Our approach can be easily extended with the same observation, i.e., by counting relevant instructions to recognize an encrypted field.
In addition to inferring the formats of received packets,
Dispatcher~\cite{caballero2009dispatcher} and P2C~\cite{kwon2015p2c} reverse engineer the formats of packets to be sent and, thus, are different from all aforementioned approaches as well as ours.

\defpar{Static Program Analysis (SPA).}
There are a few SPAs for reverse engineering protocols.
However, they either infer the formats of packets to be sent via imprecise abstract domain~\cite{lim2006extracting} 
or focus on cryptographic mechanisms~\cite{avalle2014formal}.
Our approach precisely infers the format of received packets and, thus, is different from these works.

%% file: conclusion.tex
\section{Conclusion}
In this work, we propose a static analysis that can infer protocol formats with both high precision and high recall.
Hence, the formats significantly enhance network protocol fuzzing, network traffic auditing, and network intrusion detection.
Particularly, 
our format-inference technique has helped existing protocol fuzzers find 53 zero-days with 47 assigned CVEs.
% Comparing to the state of the arts,
% a key advantage of our static analysis is that
% it does not rely on any input network packets but can infer formats with high precision, recall, and speed.
% Thus,
% we believe that
% it provides a promising alternative to the state of the arts, especially when input packets cannot ensure high coverage or are even not available.

\section{Limitations and Future Work}\label{subsec:limitations}

\noindent
Our static analysis currently is implemented for C
and does not support C++ due to the difficulty in analyzing virtual tables.
We focus on the source code and do not handle inline assembly and libraries that do not have code available.
We believe these limitations can be addressed with more engineering work.
For instance, we can use class hierarchical analysis, e.g., \cite{dean1995optimization}, to deal with virtual tables and support C++. 
We can use existing disassembly techniques, e.g., \cite{miller2019probabilistic}, to support inline assembly.
We leave them as our future work.

As discussed earlier, \tool\ employs existing techniques to deal with pointers and loops.
Thus, it inherits their limitations.
A common limitation shared by both \tool\ and all recent techniques is that the quality of inferred formats relies on the protocol implementation.
For instance, if the implementation ignores a field, the output formats will ignore it, either.
Nevertheless,
we have shown that \tool\ is promising in practice via a set of experiments.

%% file: appendix.tex
%\newpage

%\ifthenelse{\equal{\template}{ieee}}
%{% True case
%    \appendices
%}
%{% false case
%    \appendix
%}

\newpage

\appendix

\input{app-bug}

\input{app-wireshark-snort}

%\section{From Ordered AFG to Formats}\label{app:bnf}
%
%Algorithm~\ref{alg:bnf2} shows the formal algorithm that translates an ordered AFG to the BNF-style format.
%
%\begin{algorithm}[h]\footnotesize
%    \caption{Packet Format in BNF.}
%    \label{alg:bnf2}
%    \SetKwFunction{GenProduct}{bnf}
%    \SetKwFunction{VD}{VD}
%    \SetKwFunction{HD}{HD}
%    \SetKwProg{Proc}{Procedure}{}{}
%    \Proc{\GenProduct{$\mathbb{G}$}}{
%        $L\leftarrow$ new non-terminal symbol\;
%        \If {$\VD(\mathbb{G})=\mathbb{G}_a\mathbb{G}_b\dots$} {
%            $L\rightarrow\GenProduct(\mathbb{G}_a)~\GenProduct(\mathbb{G}_b)~\dots$\;
%        }
%        \ElseIf{$\HD(\mathbb{G}) = \mathbb{G}_a | \mathbb{G}_b | \dots $}{
%            $L\rightarrow\GenProduct(\mathbb{G}_a)~|~\GenProduct(\mathbb{G}_b)~|~\dots$\;
%        }
%        \Else{
%            \tcp{a single vertex containing $B[i],B[i+1],\dots, B[i+k]$} 
%            $L\rightarrow B[i]B[i+1]\dots B[i+k]$ with assertions in the vertex\;
%        }
%        \Return $L$\;
%    }
%\end{algorithm}

\input{app-proofs}

%% file: app-bug.tex
\startreviseblock
\section{Details of the Fuzzing Experiment}\label{app:bug}
\delete{\Large \textbf{A} \enskip\enskip \Large\textbf{EXAMPLE OF VULNERABILITIES}}

\noindent
Table~\ref{tab:bugdetails} shows the breakdown of the detected bugs by the bug types, the protocols, and the detectors.
The bugs we detected include integer overflow, buffer overflow, calling invalid addresses, and infinite loops.
We detected 53 bugs in total, while the baselines only detect 12 of them.

\begin{table}[h]\scriptsize
    \centering
    \caption{\revise{\# Bugs Detected by Our Fuzzer (\# by Baselines)}}\label{tab:bugdetails}
    \begin{tabular}{c|cccc}
        \toprule
        \textbf{Protocol} & Integer Overflow& Buffer Overflow & Calling Invalid Addr & Infinite Loops \\
        \midrule
        OSDP &&2 (0)&4 (0)&\\
        SSQ &&2 (0)&&\\
        BABEL &&12 (4)&&3 (0) \\
        ISIS &22 (8) & 8 (0) &&\\
        \bottomrule
    \end{tabular}
\end{table}

\dismissreviseblock

\begin{figure}[h]
    \centering
    \includegraphics[width=0.9\columnwidth]{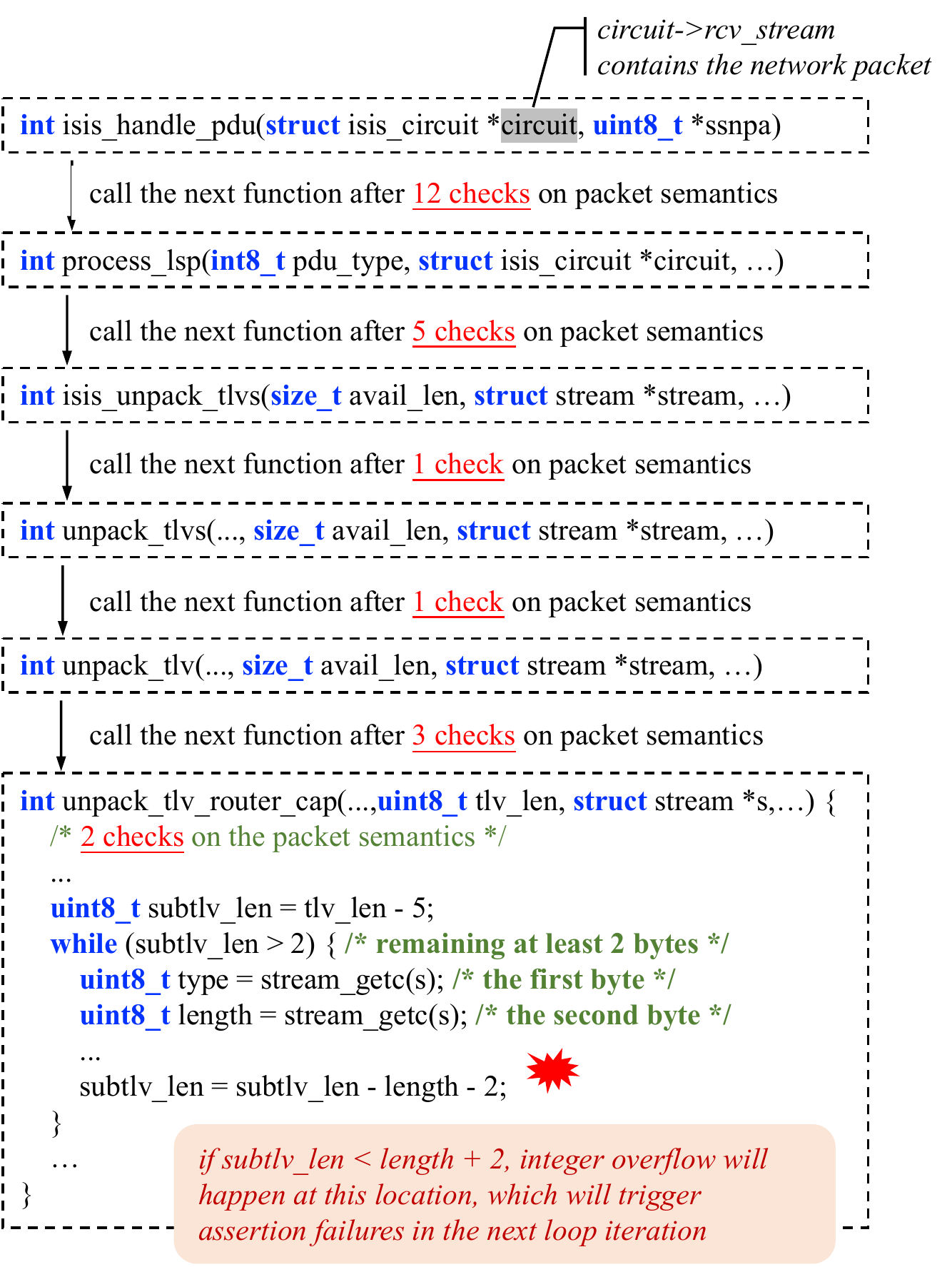}\\~
    \caption{A vulnerability in the code of the IS-IS protocol.}\label{fig:bug}
\end{figure}

Figure~\ref{fig:bug} demonstrates \replace{some details}{the details} of a vulnerability we found in \delete{an implementation of the routing protocol called} IS-IS (Intermediate System to Intermediate System)\revise{, a widely-used routing protocol}.
To perform security analysis like fuzzing, we need its format to generate valid IS-IS packets.
While it is easy to find some documents of this protocol on the internet, e.g., \cite{isis-iso,isis-techhub,isis-wiki},
all of them are written in a natural language, which cannot be directly processed by machines for automatic packet generation.
Apparently, manually translating these documents to a machine-readable formal language is labor-intensive and error-prone.
Since its implementation is available in GitHub~\cite{frrouting},
we can use our static analysis to produce its format in a formal language and, hence, facilitate the downstream automated security analysis.

The vulnerability we study here is identified by CVE-2022-26125.
Attackers may use it for DoS attacks.
This vulnerability has been fixed by the developers of this protocol. 
Thus, we do not think there are ethical concerns to discuss its details here.
As shown in Figure~\ref{fig:bug},
this vulnerability spans over at least six levels of function call, 
from the function \textit{isis\_handle\_pdu}, the entry function of the protocol parser, to the function \textit{unpack\_tlv\_router\_cap}, which parses a segment of the network packet.
Before the function call at every level,
there are at least one and up to twelve conditional statements that check if certain semantic constraints are satisfied. Ideally, these checks are sufficient to prune exploit packets.
In total, before reaching the vulnerable location,
an exploit packet needs to pass 24 checks of the semantic constraints,
which makes it hard for a fuzzer to generate such a packet via random mutation.
Hence,
a format that allows us to generate bug-triggering network packets
must precisely model such semantic constraints
and, at the same time, cannot miss packet formats that are qualified to reach the vulnerable code.
Our static analysis produces IS-IS formats with both high recall and high precision,
thereby allowing us to produce bug-triggering network packets easily. 
By contrast,
on one hand,
the format generated by NetPlier contains few semantic constraints.
Hence, packets produced based on the format usually violate the semantic constraints and, thus,
are easily filtered out by the parser.
Hence, these packets cannot execute deep program paths and are hard to trigger the vulnerability.
On the other hand,
while the format generated by Tupni is much more precise,
its recall is only 21\%, missing the format of bug-triggering packets.
Hence, the fuzzer enhanced by Tupni does not discover this vulnerability, either.

As shown in 
Figure~\ref{fig:bug},
the vulnerability happens in the function \textit{unpack\_tlv\_router\_cap},
which parses a segment of the input network packet.
In the code, the variable \textit{subtlv\_len} means the remaining bytes that have not been parsed in the segment.
The while loop can only be reached when $\textit{subtlv\_len} > 2$, i.e., at least two bytes have not been parsed.
In the loop, we read two bytes, one to the variable \textit{type} and the other to the variable \textit{length}.
The loop then parses \textit{length} bytes.  Thus, it parses $\textit{length} + 2$ bytes in total in each loop iteration. 
At the end of an iteration,
it updates the remaining bytes by subtracting the bytes parsed in the loop iteration from the variable \textit{subtlv\_len}.
Ideally, the remaining bytes, \textit{subtlv\_len}, should always decrease, until $\textit{subtlv\_len}\le 0$.
However, in the code, \textit{subtlv\_len} is an unsigned 8-bit integer and, thus, always positive.
If $\textit{subtlv\_len} < \textit{length} + 2$, e.g., $\textit{subtlv\_len} = 35$ and $\textit{length} + 2 = 36$, the subtraction will not produce an negative integer, $-1$, but a large positive integer, $255$, due to integer overflow.
This integer overflow will further lead to assertion failures in the next loop iteration as the loop expects $\textit{subtlv\_len}= 255$ remaining bytes, which do not exist.
Observe that IS-IS is a routing protocol.
This vulnerability could lead to DoS attacks 
when attackers send exploit packets to trigger the vulnerability and crash the routers.
If so, legitimate users depending on the routers will not be able to access information systems, devices, or other network resources.

%% file: app-wireshark-snort.tex
\section{Case Study via Wireshark and Snort}
\label{app:wireshark}

This appendix extends our discussion in \S\ref{sec:motivation}
to detail the attack model as well as 
comparing the effectiveness of using \tool\ and Tupni to enhance Wireshark and Snort.

\begin{figure}[t]
    \centering
    \includegraphics[width=\linewidth]{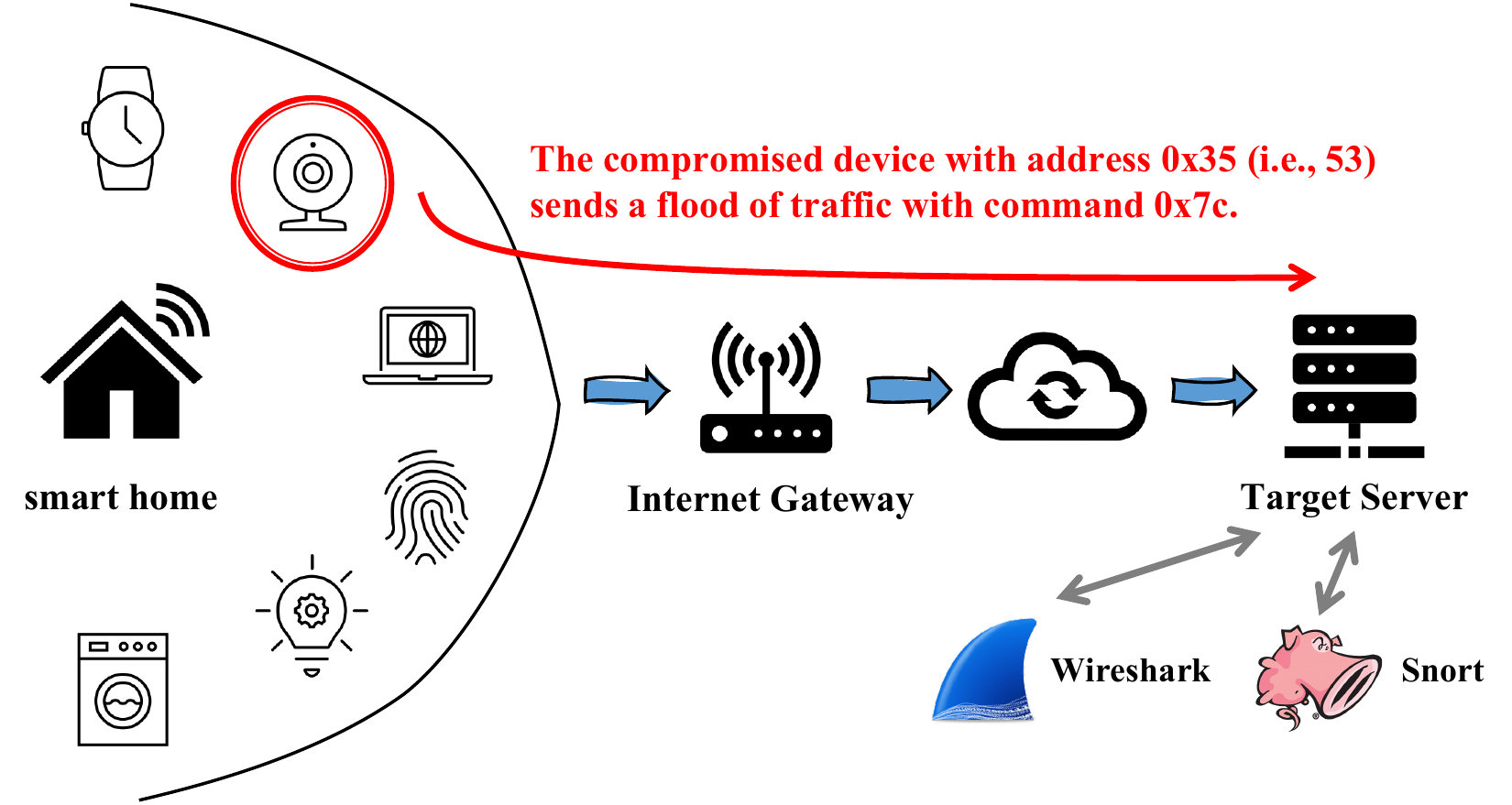}
    \caption{Attack model.}\label{fig:attackmodel}
\end{figure}

\defpar{Attack Model.}
The attack model contains a set of smart-home devices that communicate with a target using OSDP and other protocols.
One of the devices, of which the address is 0x35 as shown in Figure~\ref{fig:attackmodel}
is compromised to launch a traffic attack, i.e., send a flood of OSDP traffic with the command 0x7C
to overwhelm the target.
The target runs Wireshark and Snort, two foremost network security analyzers, to audit network traffic and detect network intrusions.
However, since the vanilla Wireshark and Snort do not support OSDP, we fail to ensure network security with them.

To support OSDP,
we use \tool\ and Tupni to infer the packet formats from the protocol's implementation.
Based on the formats,
we generate plugins to enhance Wireshark and Snort.
Ideally, the enhanced Wireshark and Snort can parse OSDP packets, dissect each packet into multiple fields, and further facilitate the analysis and detection of the traffic attack.

\begin{figure}[t]
    \centering
    \includegraphics[width=\linewidth]{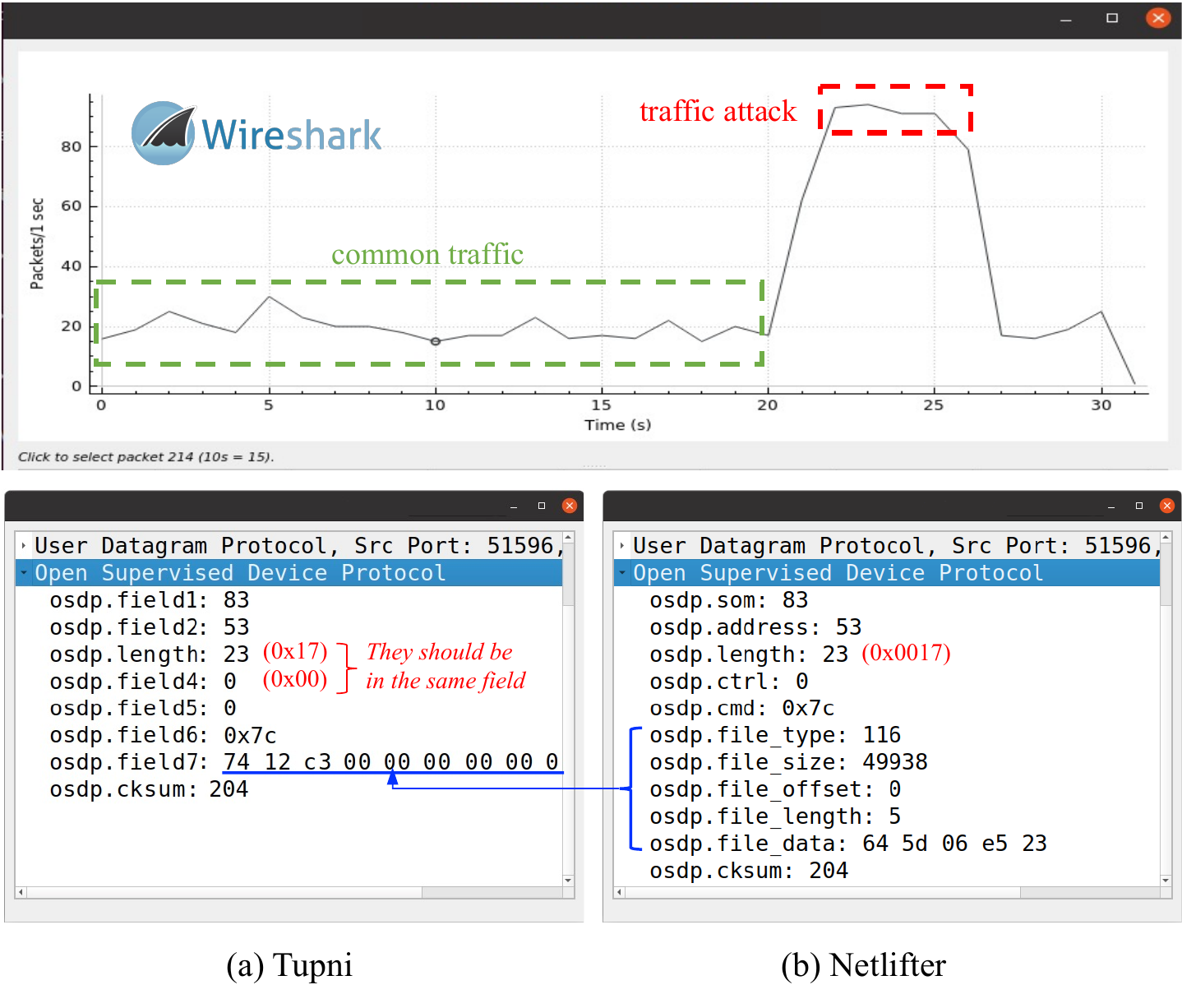}
    \caption{Screenshorts of the enhanced Wireshark.}\label{fig:wireshark-tupni}
\end{figure}

\defpar{Auditing Abnormal Traffic via Wireshark.}
We have shown in Figure~\ref{fig:motivation-2} that the vanilla Wireshark does not support OSDP. 
Thus, all OSDP packets are shown as raw bytes by Wireshark.
In this case study,
we further compare \tool-enhanced Wireshark and Tupni-enhanced Wireshark.
Figure~\ref{fig:wireshark-tupni} shows the comparison results,
which demonstrate three problems of using dynamic analysis for format inference.
First,
as a dynamic analysis,
Tupni is of relatively low coverage and misses many formats. 
As an example, in Figure~\ref{fig:wireshark-tupni}(a), 
the bytes in the 7th field are not successfully decoded
and, thus, are shown as raw bytes.
Second, it mistakenly recognizes some fields. 
For instance, the length field should contain two bytes but is split into two independent fields, i.e., \textit{length} and \textit{field4} in Figure~\ref{fig:wireshark-tupni}(a), by Tupni (note that the packet length in the ground truth should be computed as $\textit{filed4} \times 256 + \textit{length}$).
Third, it does not infer the names of most fields, making it hard to understand.
In comparison,
our static analysis is of both high precision and high coverage and, meanwhile, infers the name of all fields. 
Thus, \tool-enhanced Wireshark successfully dissects all received packets into fields, provides a proper name for each field, and, thus, effectively helps users to understand the network traffic.

\begin{figure}
    \includegraphics[width=\columnwidth]{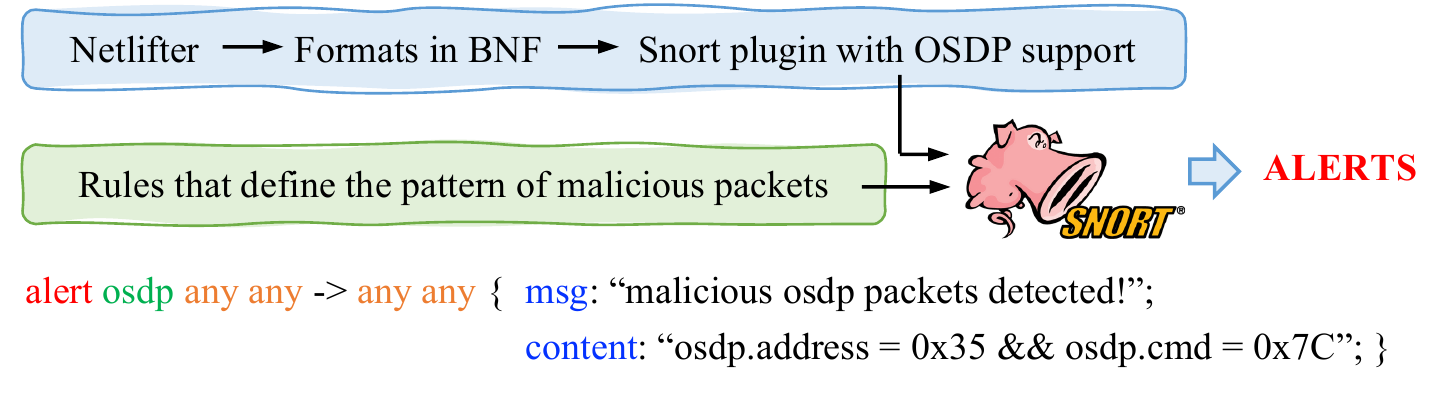}
    \caption{Rule-based intrusion detection via Snort~\cite{snort}.}
    \label{fig:motivation-3}
\end{figure}

\defpar{Detecting Intrusion via Snort.}
Snort is the foremost open-source network intrusion detection system developed by Cisco~\cite{snort}. 
It allows users to write rules to define malicious packet patterns. It then finds packets that match the patterns and generates alerts.
Figure~\ref{fig:motivation-3} shows a typical rule.
The keyword \textit{alert} indicates the action when a malicious packet is received,
\textit{any} means any source/destination ip/port of the network traffic,
\textit{msg} defines the warning message when a malicious packet is detected,
and \textit{content} defines the pattern of malicious packets, which can be determined by inspecting the attack packets (using Wireshark).
Like the Wireshark extension, we develop an extension for Snort based on the lifted protocol formats.
The extension basically parses a OSDP packet, dissects it to multiple fields, and checks the Snort rules according to the field values.

During the attack, we can use Wireshark to understand the pattern of received packets.
We then write a Snort rule as below to define the pattern of malicious packets and generate alerts for users when receiving them or directly prevent such packets.
\begin{align*}\small
    \textcolor{red}{\textup{alert}}~\textcolor{black}{\textup{osdp}}~\textcolor{orange}{\textup{any}}~\textcolor{orange}{\textup{any}}~\rightarrow&\small~\textcolor{orange}{\textup{any}}~\textcolor{orange}{\textup{any}}~\{~ \textcolor{blue}{\textup{msg}}:~\mbox{``malicious packets detected'';}\\
    &\small \textcolor{blue}{\textup{content}}:~ \mbox{``osdp.address=53~\&\&~osdp.cmd=0x7c'';}~\}
\end{align*}

When using Tupni-enhanced Wireshark, we may misunderstand the packets in the traffic attack
and define an imprecise pattern, thereby missing the chance of detecting intrusions.
As shown in Figure~\ref{fig:wireshark-tupni}(a), 
Tupni incorrectly splits the two bytes that represent length into two independent fields, \textit{length} and \textit{field4}
(note that the packet length in the ground truth should be computed as $\textit{filed4} \times 256 + \textit{length}$).
During the analysis, we find \textit{filed4} is always zero because all packets during the attack have a length less than 256.
Hence, we will define a pattern with osdp.field4 = 0 as below, which over-constrains the malicious packets:
\begin{align*}
    \small
    \textcolor{red}{\textup{alert}}~&\small\textcolor{black}{\textup{osdp}}~\textcolor{orange}{\textup{any}}~\textcolor{orange}{\textup{any}}~ \rightarrow~\textcolor{orange}{\textup{any}}~\textcolor{orange}{\textup{any}}~\{~ \textcolor{blue}{\textup{msg}}:~\textup{``malicious packets detected'';}\\
    &\small \textcolor{blue}{\textup{content}}:~ \mbox{``osdp.field2=53~\&\&~\underline{\smash{osdp.field4=0}}~\&\&~osdp.field6=0x7c'';}~\}
\end{align*}
\noindent
Next time, when attackers send packets longer than 256 bytes, the value of \textit{field4} will no longer be zero and Snort will not be able to prevent such attacks using the rule.
In our case study, when using Tupni, Snort misses over 50\% of malicious packets while \tool-enhanced Snort prevents all malicious packets.

One may think that, to prevent a traffic attack, we can simply block the IP address without looking into the OSDP packets. In fact, this does not work in many cases. For example, on one hand, we may not want to block all traffic from an IP but just block some compromised functionality, e.g., the command 0x7c. In this case, blocking the IP will overkill all functionality of the smart home devices.
On the other hand, 
the IP addresses are often dynamically allocated, blocking a single IP may not work when it is changed. However, in the smart home scenario, the OSDP address of each device can be set statically and, thus, can be reliably used.

%% file: app-proofs.tex
\section{Proofs for Theoretical Soundness \& Completeness}\label{app:proof}

\startreviseblock
Our approach infers the message formats
by building, unfolding, and reordering AFG.
Thus, the theoretical soundness and completeness of our approach is proved in the following three steps:
\begin{enumerate}
    \item Lemma~\ref{lemma:afg} proves that AFG is an equivalent representation of path constraint.
    Lemma~\ref{lemma:abstractinterpretation} shows that, given a program in our abstract language, the path constraint represented by the resulting AFG is sound and complete.
    
    \item Lemma~\ref{lemma:unfolding} states that the unfolding step does not affect the soundness and completeness of the AFG.
    
    \item Lemma~\ref{lemma:decomp}, Lemma~\ref{lemma:recursive}, and Lemma~\ref{lemma:position}
    prove three properties that hold when reordering the AFG.
    Based on these properties, Lemma~\ref{lemma:ordering} states that the reordering step does not affect the soundness and completeness of the AFG.
\end{enumerate}
\dismissreviseblock

\medskip

\textsc{Lemma}~\ref{lemma:afg}: \textit{\LemmaAFG}

\begin{proof}    
    In the proof, given any constraint $\rho$, we use $\rho_i$ to represent the conjunction of all constraints in a path of $\textsf{AFG}(\rho)$.
    
    \textbf{Base:} When a constraint $\rho$ is an atomic constraint without any connectives $\land$ or $\lor$,
    $\textsf{AFG}(\rho)$ returns a single vertex containing $\rho$. It is apparent that the lemma is correct in this trivial case.
    
    \textbf{Induction:}
    Consider two constraints, $\gamma$ and $\sigma$
    as well as their corresponding $\textsf{AFG}(\gamma)$ and $\textsf{AFG}(\sigma)$,
    which contains $m$ and $n$ paths, respectively.
    Let us assume that the lemma to prove is correct.
    That is,
    we have $\gamma \equiv \lor_{i=1}^m\gamma_i$ and  $\sigma \equiv \lor_{i=1}^n \sigma_i$.
       
    \textbf{Induction Case (1):} 
    Consider the constraint $\gamma\lor \sigma$, denoted as $\rho$. 
    We have $\textsf{AFG}(\rho) = \textsf{AFG}(\gamma) \uplus \textsf{AFG}(\sigma)$,
    which, by definition, consists of two independent subgraphs $\textsf{AFG}(\gamma)$  and $\textsf{AFG}(\sigma)$
    and, thus, contains and only contains $m+n$ paths from $\textsf{AFG}(\gamma)$and $\textsf{AFG}(\sigma)$.
    Thus,
    we have 
    \begin{equation*}
        \rho \equiv \gamma\lor \sigma \equiv  \bigvee_{i=1}^m\gamma_i \lor \bigvee_{i=1}^n \sigma_i \equiv  \bigvee_{i=1}^{m+n} \rho_i,
        \textbf{~where~}
        \rho_i = \begin{cases}
           ~\gamma_i, & i \le m \\
           ~\sigma_{i-m}, & i > m
        \end{cases} 
    \end{equation*}
    Thus, if the lemma is correct for $\gamma$ and $\sigma$, it is also correct for $\gamma\lor \sigma$.
    
    \smallskip
    
    \textbf{Induction Case (2):} 
    Consider the constraint $\gamma\land \sigma$, denoted as $\rho$. 
    we have $\textsf{AFG}(\rho) = \textsf{AFG}(\gamma) \bowtie \textsf{AFG}(\sigma)$.
    In the graph $\textsf{AFG}(\rho)$,
    all exiting vertices of the subgraph $\textsf{AFG}(\gamma)$ are connected to all entry vertices of the subgraph $\textsf{AFG}(\sigma)$.
    Hence, $\textsf{AFG}(\rho)$ contains $m\times n$ paths, and $\rho_i = \gamma_j\land \sigma_k$,
    where $1\le j\le m$ and $1\le k \le n$.
    Therefore,
    we have 
    $$\rho \equiv \gamma\land \sigma \equiv  \bigvee_{i=1}^m\gamma_i \land \bigvee_{i=1}^n \sigma_i \equiv  \bigvee_{i=1}^{m\times n} \rho_i.$$
    Thus, if the lemma is correct for $\gamma$ and $\sigma$, it is also correct for $\gamma\lor \sigma$.
    
    \textbf{Summary:} if the lemma to prove is correct for $\gamma$ and $\sigma$,
    it is also correct for $\gamma \land \sigma$ and $\gamma \lor \sigma$. Thus the lemma to prove is correct.
\end{proof}

\startreviseblock
The lemma above proves the equivalence relation between the AFG, $\textsf{AFG}(\rho)$, and the constraint $\rho$.
That is,
we can always compute the constraint it represents, i.e., $\rho$, by computing $\bigvee_i\rho_i$,
where $\rho_i$ equals the conjunction of all constraints in an AFG path.

\medskip

\textsc{Lemma}~\ref{lemma:abstractinterpretation}: \textit{\LemmaAbsInt}
\begin{proof}
    The inference rules of the abstract interpretation are shown in Figure~\ref{fig:semantics_basic}.
    For each statement in our abstract language, there is an inference rule that models its exact semantics and does not introduce any over- or under-approximation into the resulting abstract values and AFGs.
    
    For abstract values, as an example, given the abstract values $\tilde{v_2}$ and $\tilde{v_3}$ of the variables $v_2$ and $v_3$,
    the inference rule for the binary operation $v_1\leftarrow v_2\oplus v_3$ yields the abstract value  $\tilde{v_2}\oplus\tilde{v_3}$ for the variable $v_1$. This procedure does not introduce any over- or under-approximation.
    Thus, the abstract values computed by the inference rules are sound and complete.
    
    For AFGs, as an example,
    given an assertion \textbf{assert}($v$) in the code and the AFG before the assertion, e.g., $\textsf{AFG}(\rho)$, 
    the assertion rule yields a new AFG, $\textsf{AFG}(\rho)\bowtie \textsf{AFG}(\tilde{v})$, 
    which equals $\textsf{AFG}(\rho\land \tilde{v})$ by the definition of AFG.
    By Lemma~\ref{lemma:afg}, 
    the graphs $\textsf{AFG}(\rho)$, $\textsf{AFG}(\tilde{v})$, and $\textsf{AFG}(\rho\land \tilde{v})$
    represent the constraints
    $\rho$, $\tilde{v}$, and $\rho\land \tilde{v}$, respectively.
    This means that the path constraint before the assertion is $\rho$ and
    the path constraint after the assertion is $\rho\land \tilde{v}$.
    Apparently,
    Since the resulting path constraint follows the definition of path constraint
    and the abstract value $\tilde{v}$ is sound and complete, 
    this rule does not introduce any over- or under-approximation
    into the resulting path constraint $\rho\land \tilde{v}$
    and its equivalent representation $\textsf{AFG}(\rho\land \tilde{v})$.
    
    To sum up,    
    given a program written in our abstract language, 
    since each inference rule does not introduce any over- or under-approximation into the abstract values and AFGs,
    the path constraint represented by the resulting AFG is sound and complete.
\end{proof}

Note that the proof above assumes that a program is written in our abstract language,
which is loop-free. Thus, the static analysis always converges with a sound and complete result.
We discuss how we handle structures not included in the abstract language, e.g., pointers and loops, in \S\ref{subsec:impl}.
\dismissreviseblock

\medskip

\textsc{Lemma}~\ref{lemma:unfolding}: \textit{\LemmaUnfolding}

\begin{proof}
    (1)
    $\mathbb{G}_\textup{slice}$ does not miss any $\Theta_{\kappa_i}$-merged values
    because neither $\mathbb{G}_\textup{forward}$ nor $\mathbb{G}_\textup{backward}$
    misses any $\Theta_{\kappa_i}$-merged values.
    First, %
    according to the branching rule in Figure~\ref{fig:semantics_basic},
    whenever a $\Theta_{\kappa_i}$-merged value is defined,
    we have created the subgraphs, $\mathbb{G}_{\kappa_i}$
    and $\mathbb{G}_{\lnot\kappa_i}$.
    Hence, the two subgraphs can reach all uses of $\Theta_{\kappa_i}$-merged values.
    Since the graph $\mathbb{G}_\textup{forward}$ include all vertices reachable from $\mathbb{G}_{\kappa_i}$
    and $\mathbb{G}_{\lnot\kappa_i}$,
    $\mathbb{G}_\textup{forward}$ does not miss any $\Theta_{\kappa_i}$-merged values.
    Second,
    $\mathbb{G}_\textup{backward}$ does not miss any $\Theta_{\kappa_i}$-merged values
    because it is obtained by traversing the AFG from each $\Theta_{\kappa_i}$-merged value.
    
    Since $\mathbb{G}_\textup{slice}$ does not miss any $\Theta_{\kappa_i}$-merged values
    and each $\Theta_{\kappa_i}$-merged value is reachable from $\mathbb{G}_{\kappa_i}$
    and $\mathbb{G}_{\lnot\kappa_i}$,
    all $\Theta_{\kappa_i}$-merged values are replaced by either their first or second operands.
    Hence, the resulting AFG does not contain any $\Theta_{\kappa_i}$-merged value.
    
    (2)
    Recall that we copy $\mathbb{G}_\textup{slice}$ twice, one connected to $\mathbb{G}_{\kappa_i}$
    and the other connected to $\mathbb{G}_{\lnot\kappa_i}$.
    Apparently,
    this copy operation does not change the number of paths in the AFG as well as the constraint represented by each AFG path.
    
    Replacing a $\Theta_{\kappa_i}$-merged value reachable from $\mathbb{G}_{\kappa_i}$
    with its first operand also does not change the constraints represented by the AFG, due to two reasons. 
    First, by Lemma~\ref{lemma:afg}, 
    any AFG path from $\mathbb{G}_{\kappa_i}$ to a $\Theta_{\kappa_i}$-merged value
    represents the path constraint of a program path from the true branch of $\textbf{if}_{\kappa_i}$-statement.
    Second, by definition of $\Theta_{\kappa_i}$, in such a program path, the $\Theta_{\kappa_i}$-merged value equals its first operand. 
    Similarly, replacing a $\Theta_{\kappa_i}$-merged value reachable from $\mathbb{G}_{\lnot\kappa_i}$
    with its second operand does not change the constraints represented by the AFG, either.
\end{proof}

\textsc{Lemma}~\ref{lemma:decomp}: \textit{\LemmaDecomposition}

\begin{proof}
    Vertical decomposition does not change AFG and, thus, does not change the constraint represented by AFG.
    
    Given an AFG with multiple entry vertices,
    horizontal decomposition splits it into multiple subgraphs,
    each containing and only containing the vertices and edges reachable from one entry vertex.
    In other words,
    each path in a subgraph is a copy of the original AFG,
    and,
    for any path in the original AFG,
    there is a subgraph containing a copy of the path.
    This means the number of paths and the constraint in each path are not changed after horizontal decomposition.
    Hence, the constraint represented by the AFG is not changed after horizontal decomposition.
\end{proof}

\textsc{Lemma}~\ref{lemma:recursive}: \textit{\LemmaRec}

\begin{proof}
    If an AFG with multiple vertices has only one entry vertex, it at least can be vertically decomposed into two subgraphs,
    one is the entry vertex and the other is the remaining subgraph.
    Let us prove this by contradiction.
    If we cannot vertically decompose it to the entry vertex $v$ and the remaining subgraph $\mathbb{G}$,
    it must be in the following two cases.
    
    (1) The vertex $v$ is not connected to all entry vertices of $\mathbb{G}$. This means that there exists an entry vertex $v'$ in $\mathbb{G}$ that does not have any predecessors in the original graph. This further implies that the original graph has multiple entry vertices, $v$ and $v'$, which contradicts our assumption that the original AFG has only one entry vertex.
    
    (2) The vertex $v$ is connected to all entry vertices of the subgraph $\mathbb{G}$ and, meanwhile, connects to a non-entry vertex $v''$ in $\mathbb{G}$, which is reachable from an entry vertex $v'$.
    This means that there are two paths in the program with path constraints $v\land v' \land v''$ and $v\land v''$.
    We cannot write such a program in our abstract language~(Figure~\ref{fig:design_lang}).
    This is basically because, whenever we have a branching condition $v'$,
    we must have the other branching condition $\lnot v'$ that is connected to $v$.

    As discussed above, an AFG that has multiple vertices but cannot be vertically decomposed must have multiple entry vertices.
    By definition, horizontally decomposing the graph leads to multiple subgraphs, each of which starts from a single entry vertex.
    As discussed before, a subgraph containing multiple vertices but a single entry vertex can be vertically decomposed.
\end{proof}

\textsc{Lemma}~\ref{lemma:position}: \textit{\LemmaPos}

\begin{proof}
    By definition, the AFG,
    $\cdots\bowtie\textsf{AFG}(\rho_i)\bowtie\cdots\bowtie\textsf{AFG}(\rho_j)$ $\bowtie\cdots$,
    represents the constraint $\cdots \land \rho_i\land \cdots\land \rho_j\land \cdots$, denoted as $\rho$.
    
    Switching the position of $\textsf{AFG}(\rho_i)$ and $\textsf{AFG}(\rho_j)$ yields the AFG,
    $\cdots\bowtie\textsf{AFG}(\rho_j)\bowtie\cdots\bowtie\textsf{AFG}(\rho_i)\bowtie\cdots$,
    which represents the constraint $\cdots \land \rho_j\land \cdots\land \rho_i\land \cdots$, denoted as $\rho'$.
    
    By the commutative law of conjunction, $\rho$ is equivalent to $\rho'$.
    Hence, switching the position of two subgraphs in vertical decomposition does not change
    the constraint represented by the AFG.
\end{proof}

\textsc{Lemma}~\ref{lemma:ordering}: \textit{\LemmaOrdering}

\begin{proof}
    (1)
    Algorithm~\ref{alg:bnf} transforms an input AFG by decomposition or switching positions in vertical decomposition.
    By Lemma~\ref{lemma:decomp} and Lemma~\ref{lemma:position}, these operations do not change the constraint represented by the AFG. Hence, the resulting AFG of Algorithm~\ref{alg:bnf} represents an equivalent constraint as the input AFG.
    
    (2)
    The resulting AFG must be ordered, which is proved below.
    Assume that there is a path from a vertex $v_1$ to a vertex $v_2$ in the input AFG
    and, the byte index in $v_2$ is less than that in $v_1$.
    
    By Lemma~\ref{lemma:recursive},
    Algorithm~\ref{alg:bnf} can recursively split the AFG by vertical and horizontal decomposition until every subgraph after decomposition contains a single vertex.
    Whenever vertical decomposition succeeds,
    Algorithm~\ref{alg:bnf} will try to reorder the subgraphs and create the array $\mathcal{A} = [ \mathbb{G}_1, \mathbb{G}_2, \dots ]$ (Lines~3-5).
    These subgraphs contain mutually exclusive byte indices and all byte indices in $\mathbb{G}_i$ must be less than those in $\mathbb{G}_{i+k} (k\ne 0)$.
    
    (2.1) If $v_1$ and $v_2$ are in two different subgraphs $\mathbb{G}_i, \mathbb{G}_j\in \mathcal{A}$,
    it is apparent the two vertices have been correctly reordered. After reordering, the subgraphs $\mathbb{G}_i$ and $\mathbb{G}_j$ will be independently reordered. Thus, further reordering does not change the order of $v_1$ and $v_2$.
    
    (2.2) If $v_1$ and $v_2$ are in the same subgraph $\mathbb{G}_i$, 
    then $\mathbb{G}_i$ may be a merged subgraph (Lines 7-11)
    or $\mathbb{G}_i$ is obtained by vertical decomposition and, thus, cannot be vertically decomposed again (Line 13).
    No matter in which case, the subgraph $\mathbb{G}_i$ will be horizontally decomposed.
    That is to say,
    whenever $v_1$ and $v_2$ are not correctly ordered as (2.1),
    they will be in the same subgraph and the subgraph will be horizontally decomposed.

    By definition,
    continuous horizontal decomposition will let
    each subgraph contains fewer and fewer paths until
    that the two vertices $v_1$ and $v_2$ are in a single AFG path.
    Given a single AFG path, it is easy to check that    
    Algorithm~\ref{alg:bnf} will correctly reorder the two vertices as (2.1).
    
    To sum up,
    for any pair of vertices, $v_1$ and $v_2$, in an AFG path, Algorithm~\ref{alg:bnf} will correctly order them.
    Hence, the resulting AFG produced by Algorithm~\ref{alg:bnf} must be ordered.
\end{proof}